\documentclass[aps,superscriptaddress,prb,twocolumn,showpacs,letterpaper,tighten,float]{revtex4-1}
\usepackage{graphicx}
\usepackage{subfigure}
\usepackage{latexsym}
\usepackage{amsmath,amssymb}
\usepackage{bm} 
\usepackage{color}
\usepackage{stackrel}
\usepackage{epsfig}

\DeclareMathOperator{\sgn}{sgn}

\definecolor{myblue}{rgb}{.93, .93, 1}
\newcommand*\mybluebox[1]{%
\colorbox{myblue}{\hspace{1em}#1\hspace{1em}}}
\newcommand*\myFbluebox[1]{%
\fcolorbox{black}{myblue}{\hspace{1em}#1\hspace{1em}}}

\newcommand{\bsub}{\begin{subequations}}
\newcommand{\esub}{\end{subequations}}
\newcommand{\beq}{\begin{empheq}[box=\mybluebox]{align}}
\newcommand{\beqF}{\begin{empheq}[box=\myFbluebox]{align}}

\newcommand{\vex}[1]{\bm{\mathrm{#1}}}

\newcommand{\bk}{{\bf k}}
\newcommand{\br}{{\bf r}}
\newcommand{\bx}{{\bf x}}
\newcommand{\bR}{{\bf R}}

\begin{document}
	
\title{Single particle excitations in disordered Weyl fluids}
\author{J. H. Pixley}
\affiliation{Condensed Matter Theory Center and Joint Quantum Institute, Department of Physics, University of Maryland, College Park, MD 20742-4111 USA}
\author{Yang-Zhi~Chou}
\affiliation{Department of Physics and Center for Theory of Quantum Matter, University of Colorado, Boulder, CO 80309 USA}
\author{Pallab Goswami}
\affiliation{Condensed Matter Theory Center and Joint Quantum Institute, Department of Physics, University of Maryland, College Park, MD 20742-4111 USA}
\author{David A. Huse}
\affiliation{Physics Department, Princeton University, Princeton, NJ 08544 USA}
\author{Rahul Nandkishore}
\affiliation{Department of Physics and Center for Theory of Quantum Matter, University of Colorado, Boulder, CO 80309 USA}
\affiliation{Kavli Institute for Theoretical Physics, University of California, Santa Barbara, CA 93106}
\author{Leo Radzihovsky}
\affiliation{Department of Physics and Center for Theory of Quantum Matter, University of Colorado, Boulder, CO 80309 USA}
\affiliation{Kavli Institute for Theoretical Physics, University of California, Santa Barbara, CA 93106}
\affiliation{JILA, University of Colorado, Boulder, CO 80309 USA}
\author{S. Das Sarma}
\affiliation{Condensed Matter Theory Center and Joint Quantum Institute, Department of Physics, University of Maryland, College Park, MD 20742-4111 USA}
\date{\today}

\begin{abstract}
We theoretically study the single particle Green function of a three dimensional disordered Weyl semimetal using a combination of techniques. 
These include analytic $T$-matrix and renormalization group methods 
with complementary regimes of validity, and an exact numerical approach based on the kernel polynomial technique.  We show that at any nonzero disorder, Weyl excitations are not ballistic: they instead have a nonzero linewidth that for weak short-range disorder arises from non-perturbative resonant impurity scattering.  Perturbative approaches find a quantum critical point between a semimetal and a metal at a finite disorder strength, but this transition is avoided due to nonperturbative effects. At moderate disorder strength and intermediate energies the avoided quantum critical point renormalizes the scaling of single particle properties. In this regime we compute numerically the anomalous dimension of the fermion field and find $\eta= 0.13 \pm 0.04$, which agrees well with a renormalization group analysis ($\eta= 0.125$).  Our predictions can be directly tested by ARPES and STM measurements in samples dominated by neutral impurities.
\end{abstract}

\maketitle

\section{Introduction}
The exploration of Dirac and Weyl semimetals is a major activity in current
condensed matter physics, 
a subject further enriched intellectually by its deep connections to  quantum field theories and topological phenomena. While the journey began with nodal superconductors~\cite{Gorkov-1985,Nersesyan-1994}, graphene~\cite{Graphenereview2}, 
and topological insulator surface states~\cite{Bernevig2013},
in recent years the focus has shifted to three dimensional systems such as Weyl semimetals 
\cite{MurakamiWeyl, WanSavrasov, WeylBalents, WeylDai,Volovik}. These 
weakly correlated semimetallic materials
(such as TaAs Refs.~\onlinecite{Xu3-2015,Weng-2015}, NbAs Ref.~\onlinecite{Xu2-2015}, Cd$_3$As$_2$ Refs.~\onlinecite{Neupane-2014,Liu-2014,Borisenko-2014}, and Na$_3$Bi Refs.~\onlinecite{Liu2-2014,Xu-2015})
have 
bands that touch linearly at isolated points in the Brillouin zone. This gives rise to a
host of predicted exotic phenomena, including protected Fermi arc states \cite{WanSavrasov}, nonlocal quantum oscillations \cite{PKV, Moll}, and a solid state realization of the chiral anomaly \cite{SonSpivak, ParameswaranGroverAbanin, Goswami-2015, anomaly1, anomaly2}. 
This plethora of unconventional phenomena in Weyl semimetals has been established theoretically for ideal systems, and it is not a priori obvious how fragile the semimetal phase may be when effects of disorder, which must invariably be present in all real materials, are taken into account. In particular, the question of whether the Weyl semimetal is stable to weak disorder is important in this context.

{\it Disordered} Weyl semimetals present a rich and experimentally relevant challenge for condensed matter theory.  
Early work~\cite{Fradkin, Goswami-2011,Kobayashi-2014, Ominato-2014,  Syzranov-2015L, Bjorn-2014, Roy-2014,*Bitan-2016, Rex-2014, Ziegler-2015, Bjorn-2015, Syzranov-2015, Pixley-2015, Syzranov-2016, DasSarma-2015, Bera-2016,  Shapourian-2016, Pixley-2016B,Chen2015PRL,Liu2016PRL} (for a review, see Ref.~\onlinecite{SyzranovReview}) 
 suggested that weak short-range  disorder averages out (i.e. is `irrelevant' in the renormalization group sense),
and that 
the semimetallic phase (characterized by a vanishing low energy density of states (DOS)) has a non-zero regime of stability, with a quantum phase transition to a metallic phase (characterized by non-zero low energy DOS) occurring only at finite disorder.
 Physically, this suggests that in the presence of weak short-range disorder, Weyl excitations remain ballistic to asymptotically low energy, and become diffusive only  above a critical disorder strength. The quantum critical point itself presents an example of a `non-Anderson' disorder driven transition and has a rich 
 phenomenology \cite{Syzranov-2015, Pixley-2016B, SyzranovReview}. Separately, however, it has also been suggested \cite{NandkishoreHuseSondhi, Wegner-1981} that {\it non-perturbative} effects associated with rare regions give rise to a non-zero DOS for arbitrarily weak disorder, 
 calling into question the existence of a stable semimetallic phase and the disorder driven quantum phase transition.
This picture has recently been confirmed numerically~\cite{Pixley1}, with the finding that rare, low-energy, quasi-localized eigenstates contribute an exponentially small DOS at weak disorder such that the DOS remains finite to arbitrarily low disorder.

\begin{figure*}[t!]
\centering
\begin{minipage}{.45\textwidth}
\includegraphics[width=1.\linewidth]{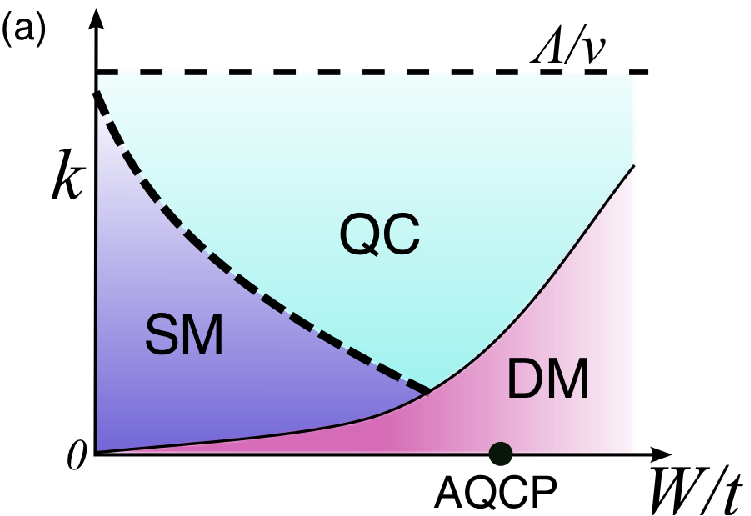}
\end{minipage}\hspace{0.5pc}
\begin{minipage}{.45\textwidth}
\includegraphics[width=0.75\linewidth,angle=-90]{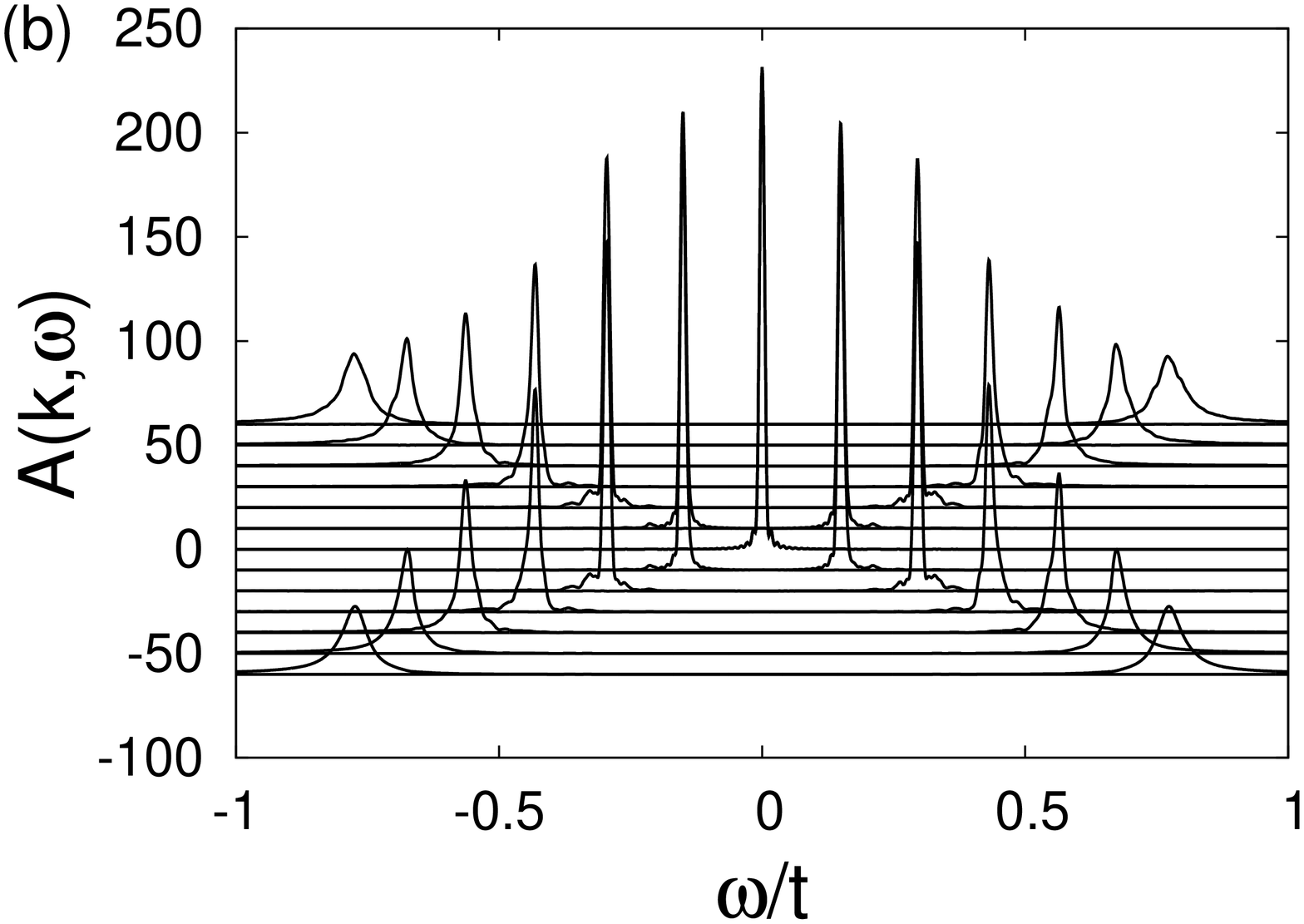}
\end{minipage}
\begin{minipage}{.45\textwidth}
\includegraphics[width=0.75\linewidth,angle=-90]{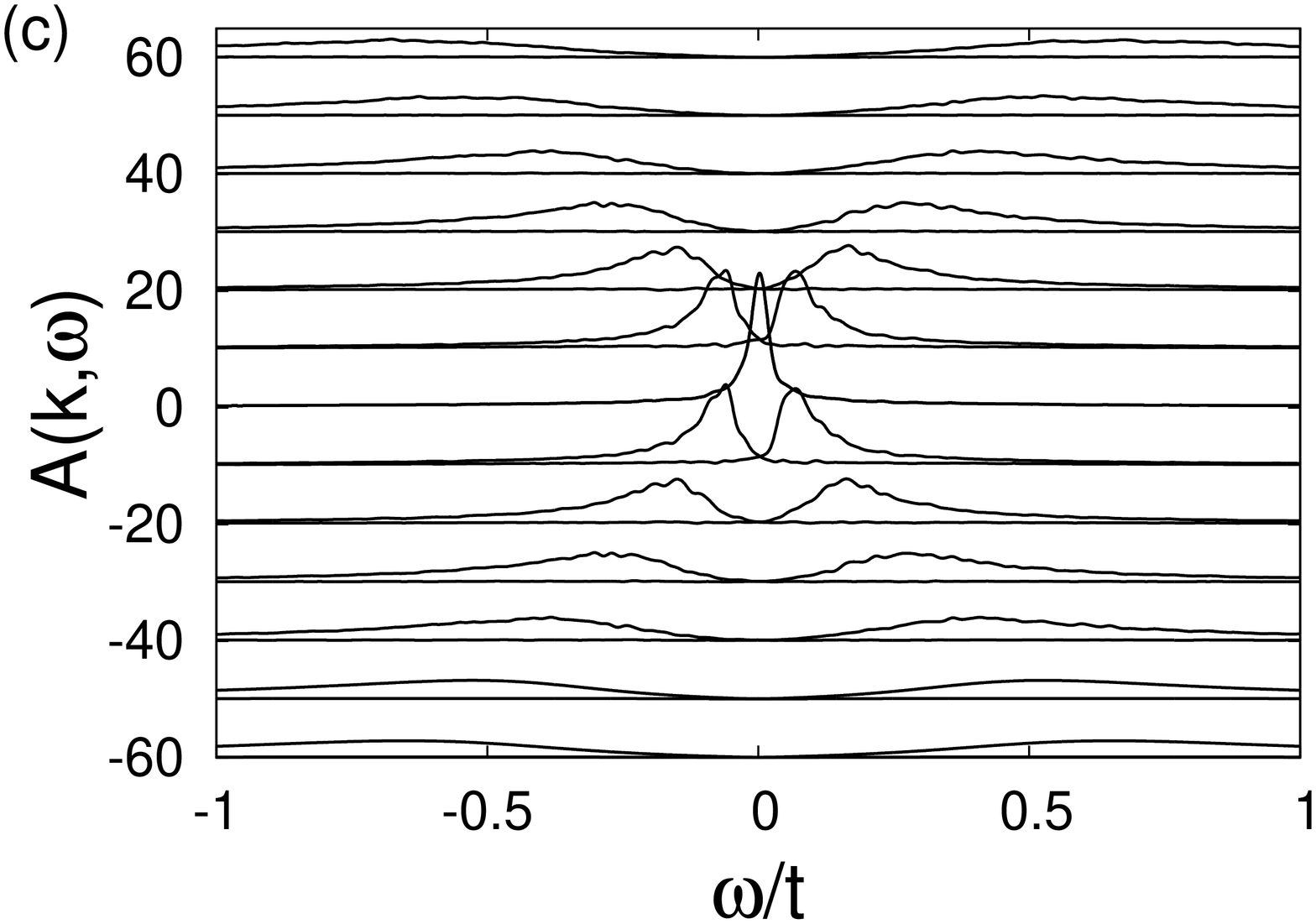}
\end{minipage}\hspace{0.5pc}
\begin{minipage}{.45\textwidth}
\includegraphics[width=0.75\linewidth,angle=-90]{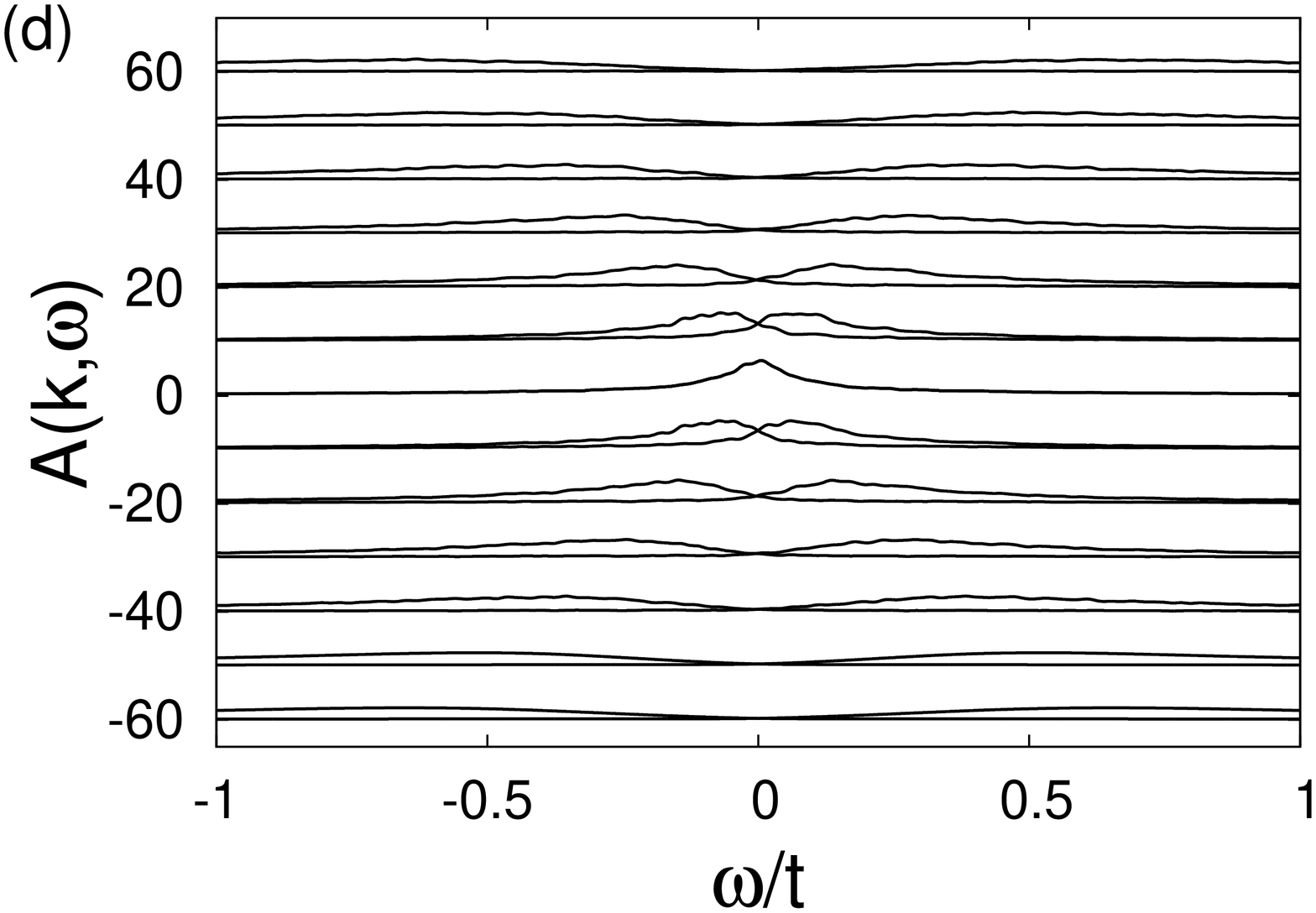}
\end{minipage} 
\caption{(color online) (a) Schematic crossover diagram for disordered Weyl semimetals as a function of momentum ($k$) and disorder ($W)$, with the diffusive metal (DM) regime at low momentum, the semimetal (SM) regime at weak disorder and intermediate momentum, and the quantum critical (QC) fan at intermediate disorder and momentum. The cut off momentum $\Lambda/v$ bounds the low energy regime.  The shape of the crossover boundaries follows from $E(\bk)$ and Refs. \cite{Pixley-2016B} and \cite{Pixley2}. 
(b-d) Electronic dispersion curves of the spectral function $A(\bk,\omega) = -G''(\bk,\omega)/\pi$ for momentum $\bk=(k,0,0)$ with $|\bk|<1$, for $L=40$ and KPM expansion order $N_C=1024$ at $W/t=0.2$ (b), 0.75 (c), and 0.9 (d). Each curve is shifted vertically by $(10kL/2\pi)$.  The quasiparticle excitations are sharp in the SM regime (b), and marginally broadened in the QC regime (c).}
\label{fig:Aw}
\end{figure*}

The possibility has also been raised that the DOS is a sum of two parts, a smooth background coming from rare regions and a non-analytic part due to the perturbative quantum critical point~\cite{Syzranov-2015}. This scenario has been ruled out by numerical calculations of derivatives of the DOS, which show that the DOS remains analytic near the Weyl node energy~\cite{Pixley1,Pixley2}.
As a result, the semimetal-metal quantum phase transition is rounded out below a small energy scale coming from non-perturbative effects, converting it into a crossover.
Nonetheless, there is a large 
 region of the phase diagram  at 
nonzero energy that is well described  by the perturbative renormalization group (RG) theory [see Fig.~\ref{fig:Aw} (a)]; thus this crossover regime has been dubbed quantum critical. However, since the non-analyticity in the DOS has been rounded out on the largest length scales, the non-Anderson disorder driven transition has  been converted to an avoided quantum critical point (AQCP).  The strength of avoidance can be tuned by suppressing non-perturbative effects~\cite{Pixley2}, but the DOS always remains analytic near the Weyl node energy.  

Notwithstanding the above progress, 
a direct probe into the nature of low energy single particle excitations is still lacking.
This information is completely contained in the single particle Green function, which 
can be directly measured in angle resolved photo-emmission spectroscopy (ARPES) and scanning tunneling microscopy (STM) experiments. In this work, we develop the theory of disordered Weyl excitations by computing the average single particle Green function $G(\bk, 
\omega)$ using a combination of analytical and numerical techniques. 
Note that we consider only {\it short range} disorder. 
We introduce an analytic $T$-matrix formalism that is controlled in the limit of dilute impurities \cite{Hirschfeld1988,Lee1993,Pepin2001,Chamon2001,Altland2002,Ostrovsky2006}, and which systematically captures the perturbative and non-perturbative effects of disorder in the weak disorder and low energy regime.
In the intermediate energy quantum critical crossover regime (where the $T$ matrix approach is invalid)
we employ a perturbative RG approach to describe the scaling of $G(\bk,\omega)$.  We also use the kernel polynomial method~\cite{Weisse-2006} (KPM) to compute $G(\bk,\omega)$ in a numerically exact fashion for sufficiently large system sizes throughout the phase diagram.  
Earlier use of the KPM method in this context~\cite{Pixley1,Pixley2} was limited to the calculation only of the DOS, which is not typically accessible to experiments directly.

Our numerical results are in excellent agreement with the analytic predictions (in each appropriate regime), and establish that the Weyl quasiparticle peaks are always broadened
for nonzero disorder strength and the quasiparticle residue remains non-zero at the Weyl node. 
Thus, we find that the disorder averaged single particle Green function is analytic near the avoided transition. We compute the renormalization of the single particle excitation spectrum within a $T$-matrix formalism and within the KPM.
In addition, we study the single parameter scaling of the Green function near the AQCP and compute the anomalous dimension using exact numerics and perturbative renormalization group methods. We make concrete predictions about the disorder induced redistribution of spectral weight, which manifests as non-trivial line shapes in $\mathrm{Im}G(\bk,\omega)$ that can be directly compared to spectroscopic experiments.   Away from very low momentum (set by the non-perturbative length scale), the quasiparticle lines are sharp (defined below) but weakly broadened in the semimetallic regime, while they are marginally broadened in the avoided quantum critical regime, as shown in Fig.~\ref{fig:Aw}.

The remainder of the paper is organized as follows: In Sec.~\ref{sec:model} we introduce the models we study and the definition of the Green function. In Sec.~\ref{sec:methods} we introduce the analytic $T$-matrix formalism, the numerical Kernel Polynomial Method, and the types of perturbative renormalization group  calculations we will compare with. In Sec.~\ref{sec:results} we present our analytic and numerical results and in Sec.~\ref{sec:conclusions} we conclude.

\section{Model and single particle Green function}
\label{sec:model}
We are interested in describing the low energy excitations of a weakly disordered Weyl fluid
characterized by a single-particle Hamiltonian
\begin{equation}
H_W = \psi^{\dag}({\bx})\left(\mp iv\bm{\sigma}\cdot \nabla + V(\bx)\right) \psi({\bx}),
\label{eqn:hamW}
\end{equation}
where $\psi^{\dag}(\bx)$ is a two component spinor that creates a Weyl fermion at position $\bx$, $v$ is the velocity, $\bm{\sigma}$ is a vector of Pauli operators, $\mp$ denotes two independent Weyl nodes, and $V(\bx)$ is a random short-range disorder potential drawn from the distribution $P[V]$. For the analytic calculations that follow we will 
work with the continuum low energy Hamiltonian in
Eq.~(\ref{eqn:hamW}), with a Gaussian distribution for $P[V]$.  

For our numerical work we 
consider the three-dimensional tight binding model from Ref.~\onlinecite{Pixley-2015}:
\begin{equation}
H_L = \sum_{\br, \mu=x,y,z}\frac{1}{2}(it\psi^{\dag}_{\br}\sigma_{\mu}\psi_{\br + \hat{\mu}} + \mathrm{h.c.}) + \sum_{\br} V(\br)\psi^{\dag}_{\br}\psi_{\br}.
\label{eqn:hamL}
\end{equation}
The hopping strength is denoted by $t$, and $\psi_{\br}$ is a two component spinor at site $\br$. We consider a cubic lattice (with a unit lattice constant) of linear size $L$ with periodic boundary conditions on each sample. In the clean limit this two band model has a dispersion $E_0(\bk)=\pm t\sqrt{\sum_{\mu}\sin(k_{\mu})^2}$ with 8 Weyl points located at the time reversal symmetric points of the Brillouin zone.  
This clean lattice Weyl model preserves time reversal symmetry but breaks inversion (${\bf k} \rightarrow -{\bf k}$). As a result, there is no anomalous Hall effect but instead the model exhibits an optical gyrotopy effect~\cite{Goswami-2015}.

KPM calculations are performed with two different disorder distributions $V(\br)$. To amplify the non-perturbative effects of rare regions~\cite{Pixley1} we sample a Gaussian distribution with zero mean and variance $W^2$.  We are able to get rid of the leading finite size effect (see Sec.~\ref{sec:dwn}) by shifting each disorder sample to satisfy $\sum_{\br}V(\br)=0$. 
 The unbounded tails of the Gaussian distribution greatly increase the probability of generating rare events due to large local fluctuations of the potential.
 To suppress the effects of rare regions and unveil the avoided quantum critical properties~\cite{Pixley2} we sample a binary distribution that takes values $\pm W$ with equal probability. The binary distribution reduces the probability to generate rare events, which makes the avoidance length scale much longer than for a Gaussian distribution~\cite{Pixley2}.  Note that disorder in the lattice model produces scattering {\it between} Weyl nodes, an effect that is ignored in the present analytic calculations.  
 Despite this difference, numerically computed rare eigenstates from the model in Eq.~(\ref{eqn:hamL}) agree quantitively with the expectations 
 from a theory of a single Weyl node, we therefore don't expect this will affect our results strongly in the weak disorder regime. However at sufficiently large disorder
the two models are distinct since the model in Eq.~(\ref{eqn:hamW}) represents two independent Weyl nodes, which do not have an Anderson localization transition~\cite{SchnyderRyu}, while the lattice model does at $W_l \approx 3.75t$ for Gaussian disorder (see Refs.~\onlinecite{Pixley-2015,Pixley1}). 
We will refer to the Hamiltonian simply as $H$.

We compute the disorder averaged retarded single particle Green function, $G(\br_i-\br_j,t) = i [\langle 0| \psi(\br_i,t)\psi^{\dag}(\br_j,0)|0 \rangle ]$, in momentum-frequency space, 
\begin{equation}
G_{\alpha\beta}(\bk,\omega) =[ \langle \bk, \alpha | \frac{1}{\omega + i \delta - H}| \bk, \beta \rangle].
\end{equation}
Here 
\begin{equation}
| \bk, \alpha \rangle = \frac{1}{\sqrt{V}}\sum_{\br}e^{-i \br \cdot \bk}\psi_{\br, \alpha}^{\dag}|0\rangle
\end{equation}
is a momentum eigenstate in the clean limit, $V=L^3$ is the volume, $\alpha,\beta$ are spinor indices, $|0\rangle$ is the single particle vacuum, $\delta \rightarrow 0^+$, and $[ \dots ]$ denotes a disorder average. 
Note that each disorder sample will have a Green function that depends on two momenta but after disorder averaging (which restores translational symmetry) these off-diagonal components vanish and we therefore only focus on $G({\bf k},\omega)$.
In all of the numerical data presented here we average over $1,000$ disorder realizations. 

We use the structure of $G(\bk,\omega)$ to extract the properties of single particle excitations. 
The models in Eqs.~(\ref{eqn:hamW}) and (\ref{eqn:hamL}) have two bands (labeled by $\pm$) that in the absence of disorder touch linearly at the Weyl points.  Diagonalizing the Green function in spinor space gives two eigenvalues that correspond to $G_{\pm}(\bk, \omega)$ for each band.  
For each disorder sample there is a set of exact eigenstates that have a non-zero overlap with the (clean) momentum eigenstates. This broadens the momentum states and  
for disorder that is not too strong, the functions $G_{\pm}(\bk, \omega)=G'_{\pm}+iG''_{\pm}$ have poles at $\omega=\pm E(\bk) - i\gamma(\bk)$,
which define the single particle dispersion $E(\bk)$. In standard many body fashion we expand near the pole to obtain
\begin{eqnarray}
G_{\pm}(\bk,\omega) &\approx& \frac{Z(\bk)}{\omega \mp E(\bk) + i \gamma(\bk)} ~,
\label{eqn:G}
\end{eqnarray}
where
\begin{eqnarray}
1/Z(\bk) &=& \partial_{\omega}G'_{\pm}(\bk, \omega)^{-1}\big|_{\omega = E(\bk)}~,
\end{eqnarray}
and
\begin{eqnarray}
\gamma(\bk) &=&Z(\bk)G''_{\pm}(\bk, E(\bk))^{-1}
\end{eqnarray}
defines the residue of the pole and the damping [or inverse lifetime $1/\tau(\bk)$] respectively. Eq.~(\ref{eqn:G}) approximates the spectral function $A_{\pm}(\bk,\omega)=-G''_{\pm}(\bk, \omega)/\pi$ as a Lorentzian centered about $E(\bk)$. 

\section{Methods}
\label{sec:methods}
\subsection{$T$-matrix}
We now describe the $T$-matrix formalism to determine the low energy excitations at very weak disorder. 
Focusing on a Gaussian distribution for the disorder potential with $W/t \ll 1$, it is natural to expect that the perturbative corrections will almost be exact. However, due to the unbounded tails of $P[V]$ it is possible that at some site $\bR$, there arises resonant scattering which is inherently non-perturbative. 
The most natural way to describe such a process is to consider the impurities to be dilute (with density $n_{\mathrm{imp}}b^3 \ll 1$, where $b$ is the radius of the impurity well) and solve for the single scattering event non-perturbatively (i.e. including resonance). This is achieved by considering the potential,
$
V(\vex{x})=\sum_{j=1}^{N_{\mathrm{imp}}}\lambda_j \Theta(b-|\vex{x}-\bR_j|),
$
which describes $N_{\mathrm{imp}}(=n_{\mathrm{imp}}L^3)$ randomly distributed spherically symmetric square-well potentials with a fixed width $b$ and a strength $\lambda_j$ that is randomly distributed following 
a gaussian distribution with zero mean and variance $\tilde{W}^2$. The relation between $W$ and $\tilde{W}$ is then given by $W^2=n_{\text{imp}}b^6\left(\frac{4\pi}{3}\right)^2\tilde{W}^2$.
The geometric factor $b^6\left(\frac{4\pi}{3}\right)^2$ can change for different shapes of the potential.
Approximating the impurities as square wells is for convenience, and is not expected to affect the physics \cite{NandkishoreHuseSondhi}. 

The disorder averaged (both impurity position and strength) Green function is given by
\begin{align}
\nonumber\left[G_{\alpha\beta}\left(\bk,\omega\right)\right]
\approx & G^{(0)}_{\alpha\beta}\left(\bk,\omega\right)\\
\label{Eq:dis_ave_G}&+n_{\text{imp}}G^{(0)}_{\alpha\delta}\left(\bk,\omega\right)\!
\left[\Gamma^{(\lambda)}_{\delta \delta'}(\omega;\vex{k},\vex{k})\right]\!
G^{(0)}_{\delta'\beta}\left(\bk,\omega\right),
\end{align}
where we have used the Einstein summation convention. $G^{(0)}$ denotes the Green function in the clean limit; $\Gamma^{(\lambda)}$ is the vertex function due to position averaged disorder scattering with a fixed strength $\lambda$. Again, $[\dots]$ indicates average over $\lambda$.
One can show that $\Gamma^{(\lambda)}$ in the dilute limit ($n_{\text{imp}}b^3\ll 1$) is identical to the T-matrix of the single impurity well with the strength $\lambda$ (see Ref.~\onlinecite{Mahan2000}). The disorder averaged Green function in the dilute limit only constitutes the diagonal part (in momentum) of the vertex function.

The disorder averaged Green function satisfies the Dyson equation
$
G_{\alpha\beta}\left(\bk,\omega\right)^{-1}=
G^{(0)}_{\alpha\beta}\left(\bk,\omega\right)^{-1}-
\Sigma_{\alpha\beta}(\bk,\omega)
$.
In the limit of dilute impurities we find that the self energy is momentum independent and given by 
\begin{eqnarray}
\Sigma_{\alpha\beta}(E)=\delta_{\alpha\beta}n_{\text{imp}} \int d\lambda \tilde P[\lambda] T^{(\lambda)}(\vex{k},\vex{k})\big|_{|\vex{k}|=E/(\hbar v)},
\end{eqnarray}
where $T^{(\lambda)}$ is the $T$-matrix for scattering off a single impurity well with strength $\lambda$. 
We solve for the $T$-matrix analytically via quantum mechanical scattering theory focusing on a single potential well.
The phase shift ($\delta_j$ is the phase shift for total angular momentum $j$) of the scattering problem is known exactly~\cite{NandkishoreHuseSondhi}
\begin{align}\label{Eq:phase_shift}
\tan \delta_{j}
=\frac{\sgn(q/k)J_{j}(|k|b)J_{j+1}(|q|b)-J_{j}(|q|b)J_{j+1}(|k|b)}{\sgn(q/k)Y_{j}(|k|b)J_{j+1}(|q|b)-J_{j}(|q|b)Y_{j+1}(|k|b)},
\end{align}
where $q=k-\frac{\lambda}{\hbar v}$, $J_n$ ($Y_n$) is the Bessel function of the first (second) kind with order $n$.

The diagonal $T$-matrix for the Weyl problem at hand can be expressed in terms of the phase shift via 
\begin{align}
T^{(\lambda)}(\vex{k},\vex{k})=&-\frac{2\pi \hbar v}{k}f^{(\lambda)}(\vex{k},\vex{k}),\\
f^{(\lambda)}(\vex{k},\vex{k})=&\sum_{j}(2j+1)\frac{e^{i2\delta_j}-1}{2ik},
\end{align} 
where $f^{(\lambda)}(\vex{k},\vex{k}')$ is the scattering amplitude of the scattering problem from a single impurity with a strength $\lambda$.

In the low-energy long wavelength limit, we focus on the $j=1/2$ sector 
as $j \geq 3/2$-sectors contribute higher powers of $|\mathbf{k}|$. One can see this by considering the small $|\vex{k}|b$ expansion of the $T$-matrix in the $j=1/2$ sector gives
\begin{align}
\nonumber T^{(\lambda)}_{1/2}(\vex{k},\vex{k})=&\lambda \left[\tilde{U}_0+O(|\vex{k}|^2)\right]\\
\label{Eq:Tmatrix_1}&+\lambda^2\tilde{U}_0^2\left[-\frac{2 \left(\hbar v k\right)}{5 b \hbar^2 v^2}-i\frac{\left(\hbar vk\right)^2}{4\pi\hbar^3v^3}\right]+\dots,
\end{align}
where $\tilde{U}_0=\frac{4\pi}{3}b^3$ is the Fourier transform of $\Theta(b-|\vex{x}|)$ at $\vex{k}=0$.
The leading term in Eq.~(\ref{Eq:Tmatrix_1}) matches the impurity potential. The second line in Eq.~(\ref{Eq:Tmatrix_1}) gives the correct Born approximation contribution.
For the finite momentum dependence of the single particle properties we consider contributions from both $j=1/2$ and $j=3/2$ sectors.
 At weak disorder this method captures both the perturbative and non-perturbative effects of disorder, and allows us to calculate the parameters $E(\bk)$, $Z(\bk)$ and $\gamma(\bk)$ in Eq.~(\ref{eqn:G}) from a reduced problem for a single impurity.
 
 Finally, let us discuss the regime of validity of the $T$-matrix calculation. Firstly and most obviously, this calculation technique ignores coherent scattering between different impurities. As such it is only well controlled in the limit of {\it dilute} impurities $n_{imp} b^3 \ll 1$. Additionally, the $T$ matrix is evaluated using phase shifts in the $j=1/2$ and $j=3/2$ channels only. 
 This provides an accurate estimate of the $T$ matrix in the long wavelength regime, since higher angular momentum channels enter with larger powers of $k$. However, this also implies that the method will fail at large momenta.
 Finally, and most subtly, the phase shifts are extracted from the asymptotic forms of the wavefunctions for a system with a single impurity. This only works if the wavefunctions have reached their asymptotic form on the length scale of the typical inter-impurity spacing. As pointed out in Ref.~\onlinecite{NandkishoreHuseSondhi}, on the {\it lowest} energy scales, this approximation fails, and hybridization of wavefunctions centered on different impurities  (i.e. the hybridization of multiple power-law localized states) must be taken into account. 
 However, (as also pointed out in Ref.~\onlinecite{NandkishoreHuseSondhi}), this last approximation only fails on energy scales $E < \nu_0$, where $\nu_0$ is the low energy density of states, whereas non-perturbative resonant scatterings from a single impurity already dominate the self energy for $E < \sqrt{\nu_0}$. There is thus a parametrically broad regime of energies $\nu_0 < E < \sqrt{\nu_0}$ where the single impurity $T$-matrix approach is valid.

\subsection{KPM}
The KPM expands the imaginary part of $G(\bk, \omega)$ (denoted as $G''$) in terms of Chebyshev polynomials [$T_n(x)$] to an order $N_C$ and we use the Lorentz kernel to filter out Gibbs oscillations. The real part of the Green function is obtain using the Kramers-Kronig relation~\cite{Weisse-2006}. This yields the KPM expression for the Green function
\begin{eqnarray}
G_{\alpha\beta}(\bk,\omega) &=& \Big[-\frac{i}{a\sqrt{1-\tilde{\omega}^2}}\big(\mu_0(\bk, \alpha,\beta)g_0 
\nonumber
\\
&+& 
2\sum_{n=1}^{N_C-1} \mu_n(\bk, \alpha,\beta)g_n e^{-i n\arccos\tilde{\omega}}  \big)\Big],
\end{eqnarray}
where $g_n$ denotes the Kernel, $\tilde{\omega}=(\omega-b)/a$, where $a$ is the half-bandwidth and $b$ is half of the band asymmetry. The coefficients of the expansion are given by
\begin{equation}
\mu_n(\bk, \alpha,\beta) = \langle  \alpha, \bk | T_n(\tilde{H}) |\bk, \beta \rangle,
\end{equation}
and $\tilde{H}=(H-b)/a$ is the rescaled Hamiltonian.
The Lorentz kernel broadens each Dirac-delta function in the spectral function $A(\bk,\omega)=-G''(\bk,\omega)/\pi$ into a Lorentzian~\cite{Weisse-2006} of width $\lambda D/N_C$ (for a bandwidth $D$), and $\lambda$ controls both the width of the Lorentzian and the strength of Gibbs oscillations due to truncating the series. Here, we work with $\lambda=0.5$ so that we can accurately compute the intrinsic broadening due to disorder.
\begin{figure}[h!]
\centering
\begin{minipage}{.45\textwidth}
\includegraphics[width=0.75\linewidth, angle=-90]{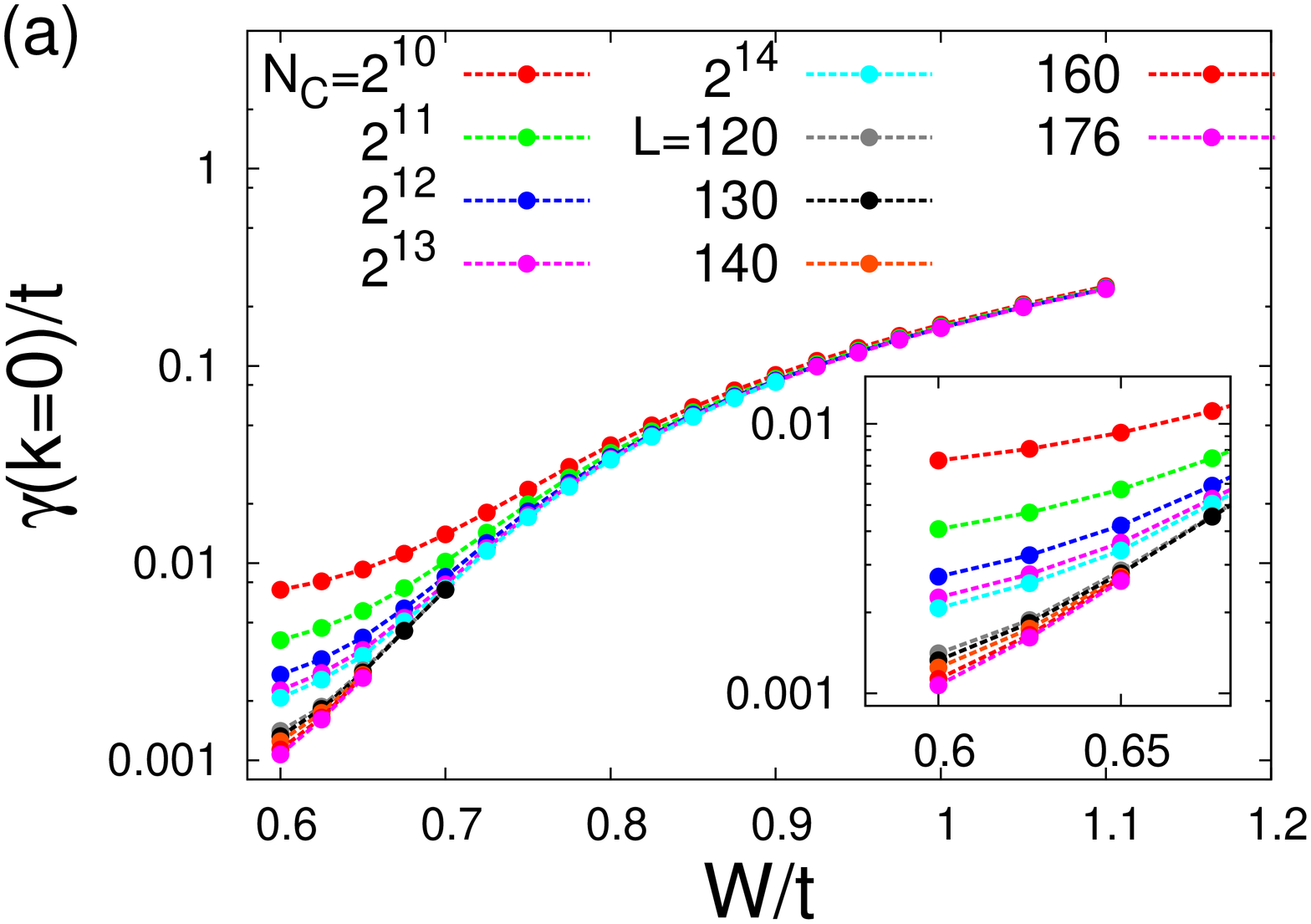}
\end{minipage}\hspace{0.5pc}
\centering
\begin{minipage}{.45\textwidth}
\includegraphics[width=0.75\linewidth,angle=-90]{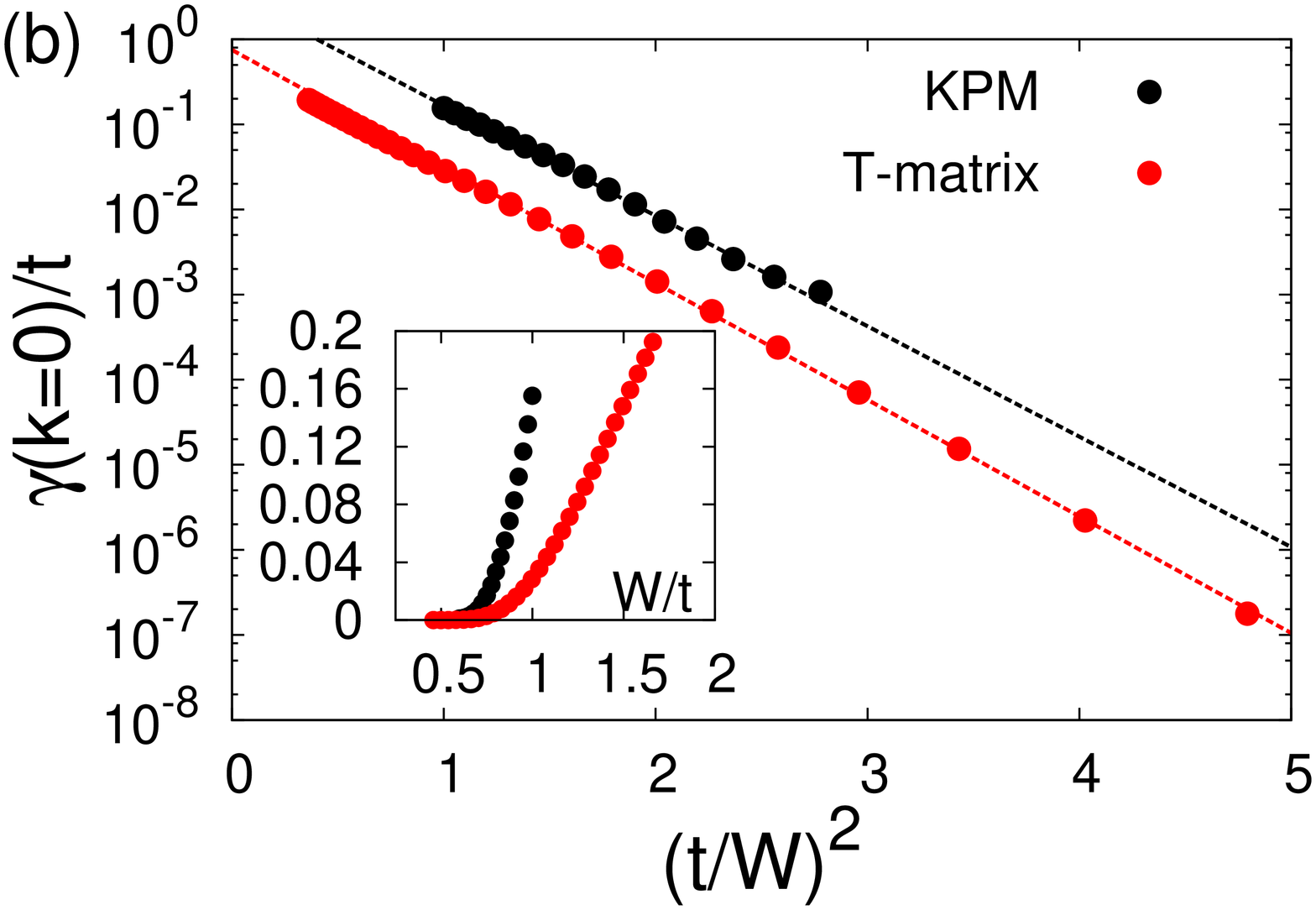}
\end{minipage}
\caption{(color online) 
(a) The convergence of $\gamma(0)$ with $L$ and $N_C$ for a gaussian disorder distribution. Data for different $N_C$ is with $L=80$ and data with $L \ge 120$ is for $N_C=2^{14}$. $\gamma(0)$ is well converged for $W \ge 0.625 t$, $L=176$, and $N_C=16384$, whereas for smaller values of $W$ we are not able to clearly discern between the artificial KPM broadening or the intrinsic broadening due to disorder. (Inset) Zoomed in on the low $W$ region displaying the strong finite $L$ and $N_C$ effects.
(b) The damping $\gamma(0)$ of the disordered Weyl point on a log scale computed from the $T$-matrix in the dilute limit for $H_W$ and the KPM for a gaussian disorder distribution for $H_L$ (converged in $L$ and $N_C$) versus $1/W^2$.  
Dashed lines are fits to the rare region form in Eq.~(\ref{eqn:rr}). For the $T$-matrix, we consider $n_{\text{imp}}b^3=0.039$ to get the best comparison with the KPM results, and $t$ is replaced by the unit of energy $E_0=\hbar v/b$. The offset arises because although the leading $W$ dependence of $\gamma(0)$ is universal, the pre-exponential factor is sensitive to the difference between Eqs.~(\ref{eqn:hamW}) and (\ref{eqn:hamL}). (Inset) $\gamma(0)$ versus $W$ on a linear scale.}
\label{fig:k=0}
\end{figure}

\subsection{Modified RG scheme}
There are different schemes for controlling the perturbative RG calculations for disordered Dirac and Weyl fermions.
For a massless Dirac system in $d$ spatial dimensions, under the scale transformation $x \to xe^l$, the disorder coupling for Gaussian white noise 
disorder scales as $\Delta(l)=\Delta(0) e^{(2-d)l}$. Therefore, for a three dimensional system we can carry out a $d=2+\epsilon$ expansion and set 
$\epsilon=1$ at the end of calculations. 
Such an analysis at one loop order predicts the correlation length and dynamic scaling exponents $\nu=1/\epsilon=1$ and $z=1+\epsilon/2=3/2$, which agrees well with high accuracy numerical calculations~\cite{Pixley2}.
Within this scheme, the one loop fermion self energy is independent of momentum and only depends on frequency and 
one finds $\eta=0$ at one loop order.

For any dimension $d$, disorder potentials with a $1/r^2$ correlation act as a marginal perturbation for Dirac fermions. Therefore perturbative calculations can also be controlled by varying the range of the probability distribution. This can be seen from the following arguments by considering a generalized power law distribution for the random potential
 \begin{eqnarray}
 \langle V(\mathbf{x}) V(\mathbf{y}) \rangle \sim \frac{\Delta}{ |\mathbf{x}-\mathbf{y}|^{d-\alpha}},
 \label{eqn:pot}
 \end{eqnarray}
  and $\alpha=0$ corresponds to the Gaussian white noise distribution~\cite{Goswami-2016}. After fixing the spatial dimensionality $d=3$, the disorder coupling scales as $\Delta(l)=\Delta(0) e^{(\alpha-1)l}$. Therefore, we can control the perturbative RG analysis with respect to the marginal case $\alpha=1$ by choosing $\alpha=1-\epsilon$ and setting $\epsilon=1$ at the end of our calculations. 
One loop analysis within this scheme also predicts the same correlation length and dynamic scaling exponents $\nu=1/\epsilon=1$ and $z=1+\epsilon/2=3/2$ as found within the $d=2+\epsilon$ expansion but in contrast predicts a nontrivial  value of $\eta = 1/8$.
 We will use these two different RG schemes below when comparing to the KPM data.

\section{Results}
\label{sec:results}
\begin{figure}[h!]
\centering
\begin{minipage}{.45\textwidth}
\includegraphics[width=0.7\linewidth,angle=-90]{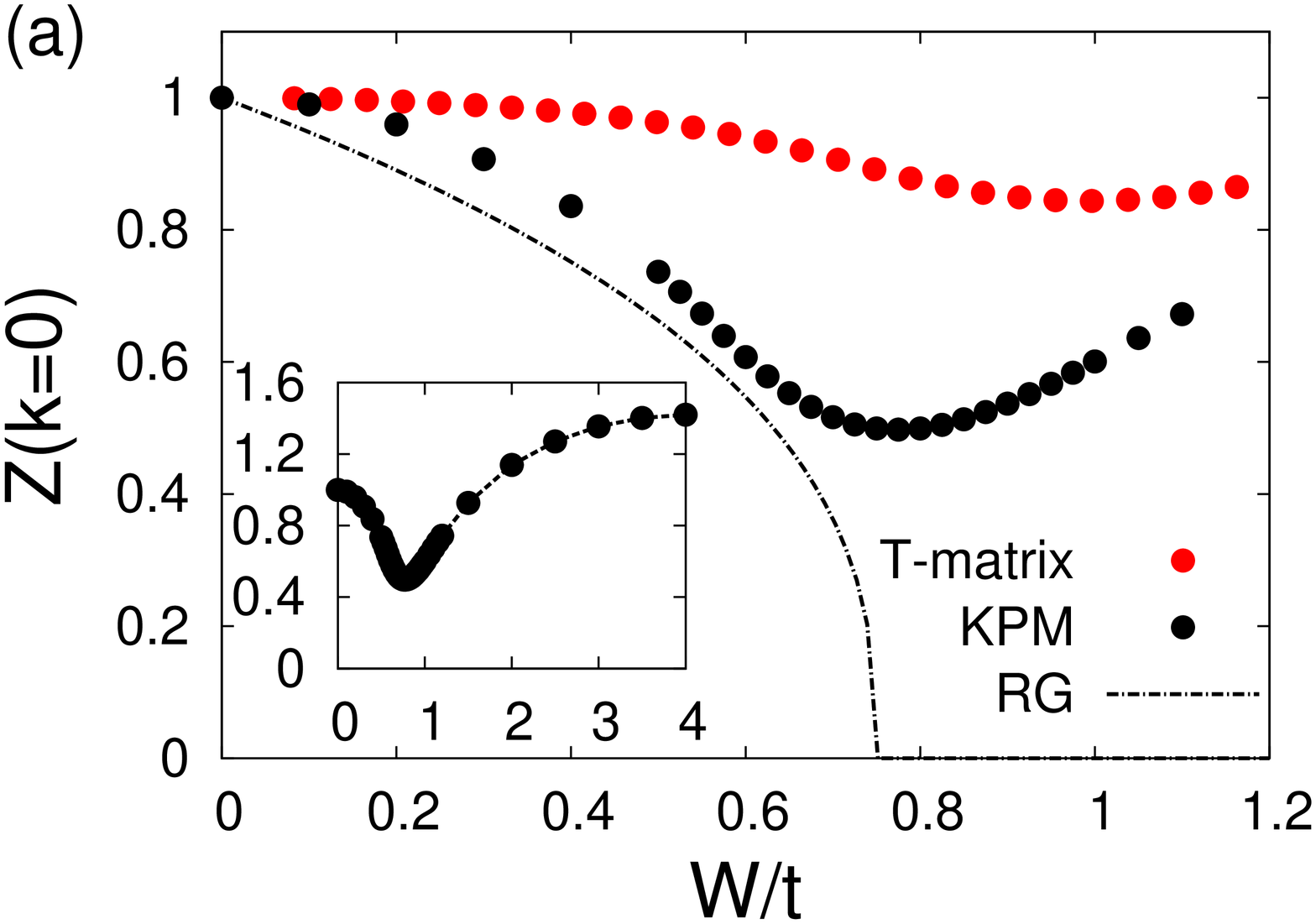}
\end{minipage}
\newline
\centering
\begin{minipage}{.45\textwidth}
\includegraphics[width=0.7\linewidth,angle=-90]{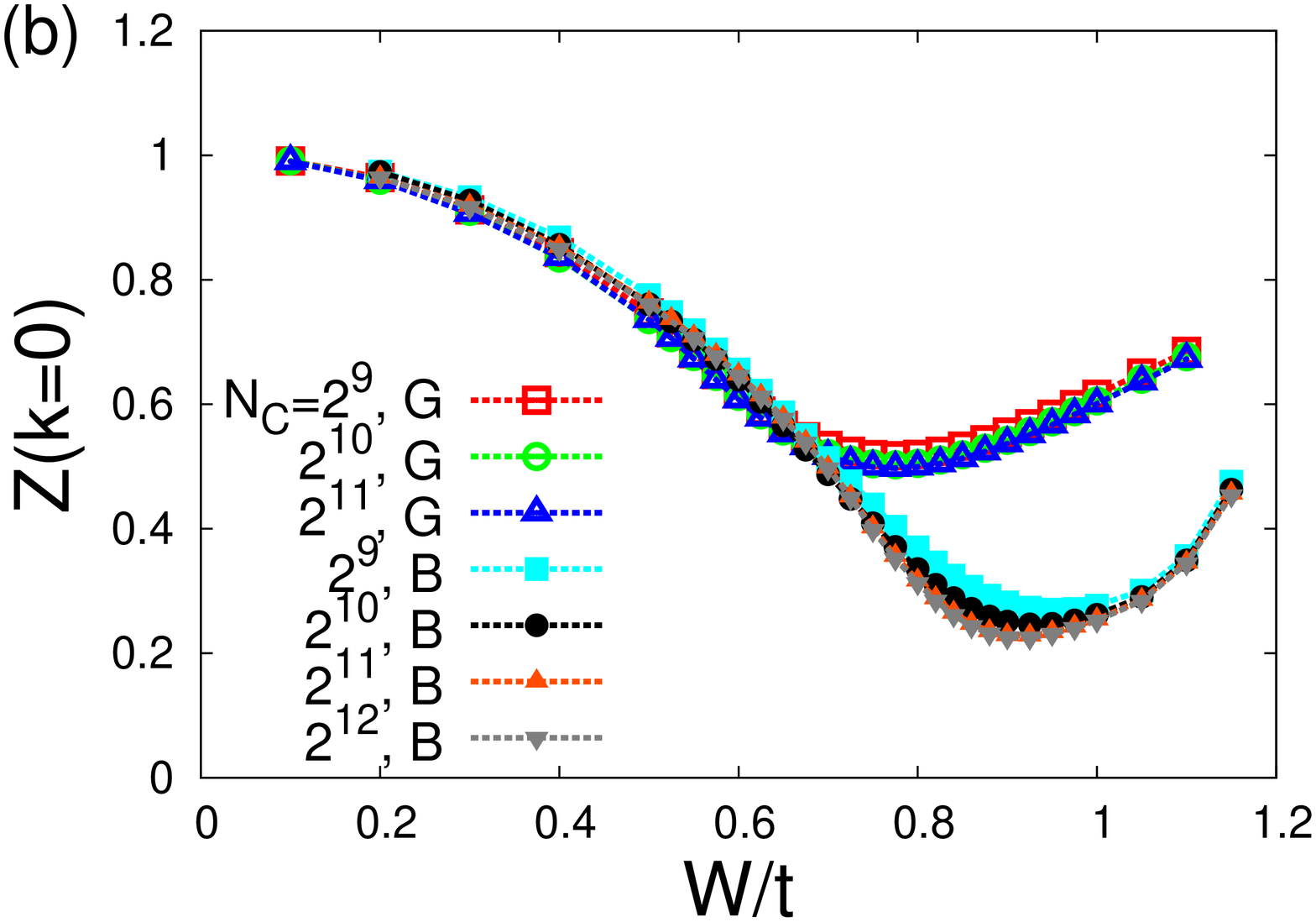}
\end{minipage}
\caption{(color online) (a) The quasiparticle residue at the Weyl node $Z(\bk=0)$ as a function of the disorder strength computed within the three methods we use in this manuscript. The dashed line shows the qualitative expectation of the RG, $Z(0)\sim (1-W/W_c)^{\nu(z-1- \eta)}$, with $\nu(z-1- \eta)=0.375$~\cite{Goswami-2016}. We find $Z(0)$ has a minimum near the AQCP ($W_c/t\approx 0.75$ for Gaussian) but does not go to zero. (Inset) $Z(0)$ up to large disorder strengths passing through the Anderson localization transition ($W_l\approx 3.75t$, Ref.~\cite{Pixley1}) at $E=0$. (b) The quasiparticle residue at $\bk=0$ for Gaussian (G) and binary (B) disorder as a function of disorder for various expansion orders. We find $Z(0)$ is well converged and has a minimum near the avoided quantum critical point, with a value that is dictated by the strength of avoidance. }
\label{fig:zk=0}
\end{figure}

We have computed the single particle Green function using KPM, yielding the 
results shown in Figs.~\ref{fig:Aw} (b), (c), and (d). 
We refer to the disorder strength relative to the location of the AQCP in the crossover diagram of the models, see Fig.~\ref{fig:Aw} (a). The semimetal (SM) regime occurs at finite but low energy and weak disorder ($W<W_c \approx 0.75t$ for Gaussian disorder), whereas the quantum critical (QC) scaling regime exists at nonzero energy for moderate disorder strengths ($W \approx W_c$), and the diffusive metal (DM) regime occurs for all energies with a crossover boundary that grows with increasing $W$ (here large disorder refers to $W>W_c$). 
At weak disorder and low momentum the spectral function is well described by a Lorentzian shape, whereas at large momentum the spectral function is broad and asymmetric such that the approximation in Eq.~(\ref{eqn:G}) is no longer valid. 
 Despite the disorder broadening of the momentum eigenstates, the average spectral function satisfies the sum rule 
 \begin{eqnarray}
\frac{1}{2L^3} \int d\omega \sum_{\bk,p=\pm}A_p(\bk,\omega)=1,
 \end{eqnarray}
  which also restricts $Z(\bk)$. This implies that if $Z(\bk)<1$ for some $\bk$, then  $Z(\bk')>1$ at some other $\bk'$ is required. This is in contrast to interacting systems, where the ``missing'' quasiparticle residue goes into incoherent spectral weight due to inelastic scattering and $Z(\bk) \le 1$ for all $\bk$ is possible.

 For the quasiparticle excitation to be ``sharp'' the two spectral functions centered about $\pm E(\bk)$ need to have very little overlap.
Therefore, \emph{sharp} quasiparticle excitations are only well defined for $E(\bk) \gg \gamma(\bk)$. 
Note that the excitations near $\bk = \pi/2(\pm 1,\pm 1,\pm 1)$ are at the upper and lower band edges, and become Anderson localized for a weak amount of disorder. 
By considering $\bk = (k,0,0)$ and $W/t \le 1$, we ensure that the single particle mobility edge never crosses $E(\bk)$ and all the excitations we discuss in this work are delocalized.

\subsection{Disordered Weyl node}
\label{sec:dwn}
\begin{figure}[h!]
	\centering
		\begin{minipage}{.45\textwidth}
		\includegraphics[width=0.75\linewidth, angle=-90]{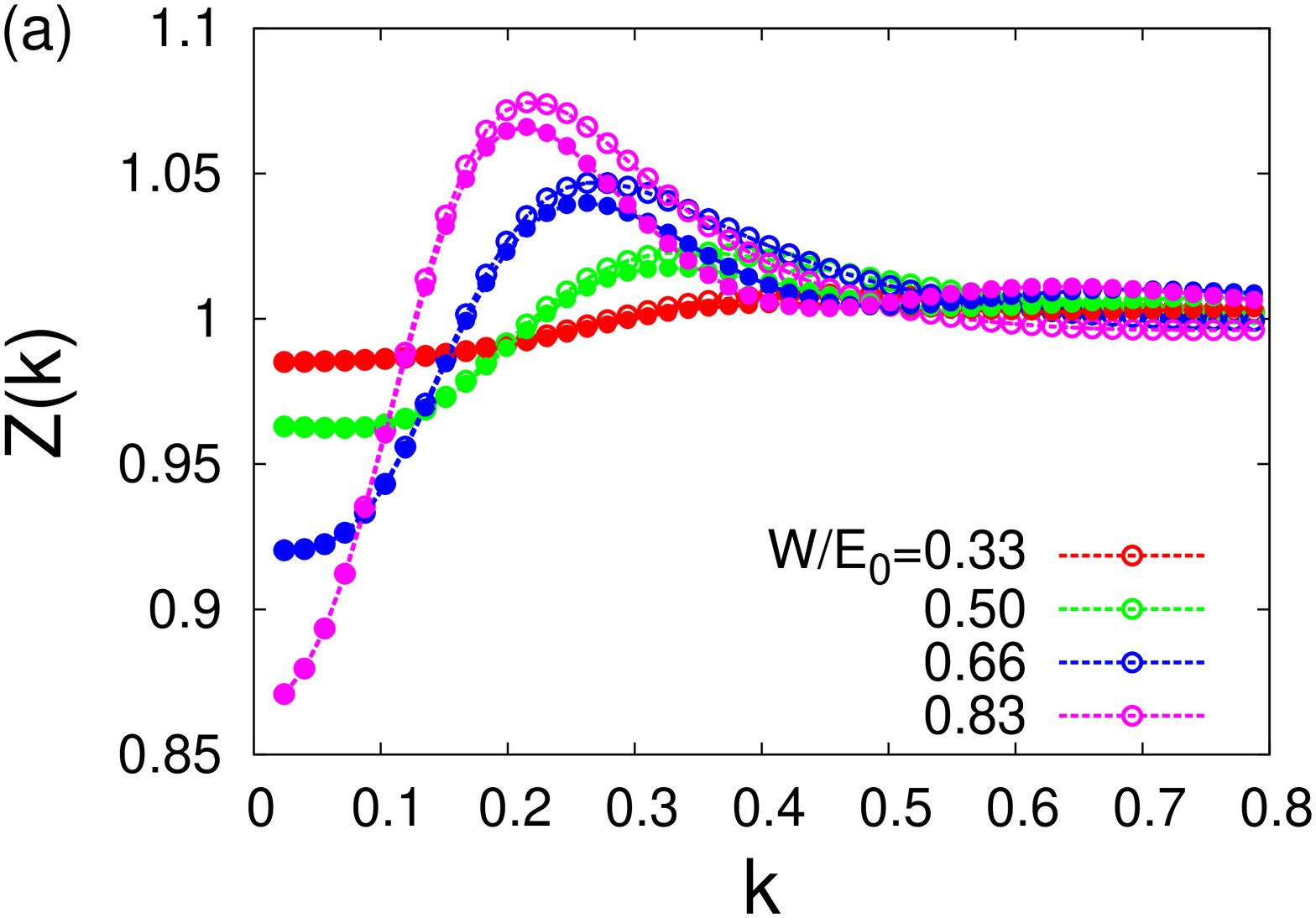}
	\end{minipage}\hspace{0.5pc}
	\centering
	\begin{minipage}{.45\textwidth}
		\includegraphics[width=0.75\linewidth, angle=-90]{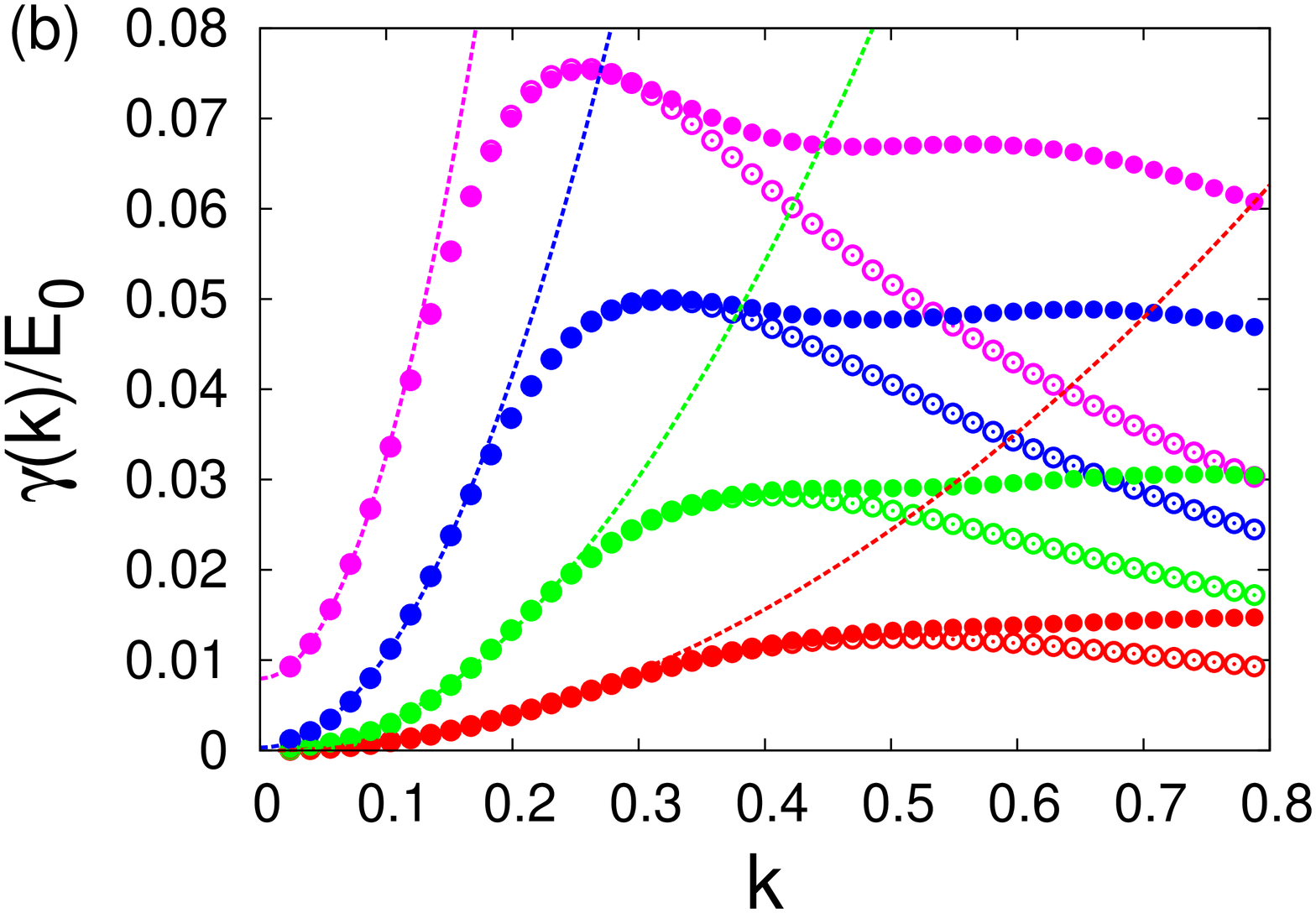}
	\end{minipage}
	\caption{(color online) Momentum dependence of the quasiparticle residue (a) and the damping (b)  from the $T$-matrix formalism. We set $n_{\text{imp}}b^3=0.34$ among all the data. The legend is shared across the two plots, the unit of energy is $E_0=\hbar v/b$, and open symbols denote the contributions from only the $j=1/2$ momentum sector and filled symbols denote the contributions from both $j=1/2$ and $j=3/2$ sectors. Dashed lines are a fit to $\gamma({\bf k})=\gamma(0)+a|\bk|^2$. 
	The quasiparticle residue becomes larger than one for some finite $k$. 
	The $j=3/2$ contribution plays an important role for $k>0.3$.}
	\label{fig:Tmatrix_z}
\end{figure}

\begin{figure*}
\centering
\begin{minipage}{.32\textwidth}
\includegraphics[width=0.75\linewidth,angle=-90]{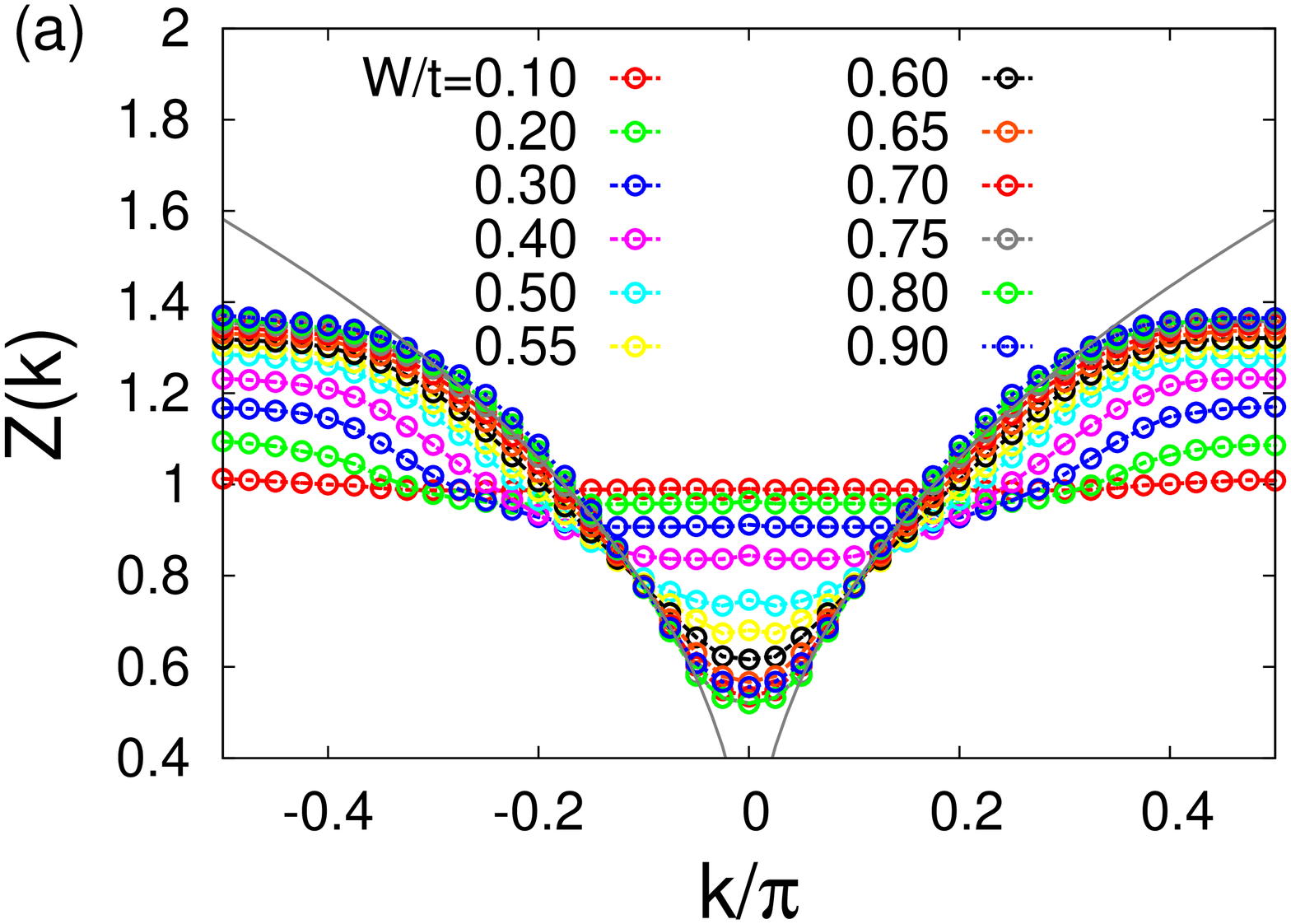}
\end{minipage}\hspace{0.5pc}
\begin{minipage}{.32\textwidth}
\includegraphics[width=0.75\linewidth,angle=-90]{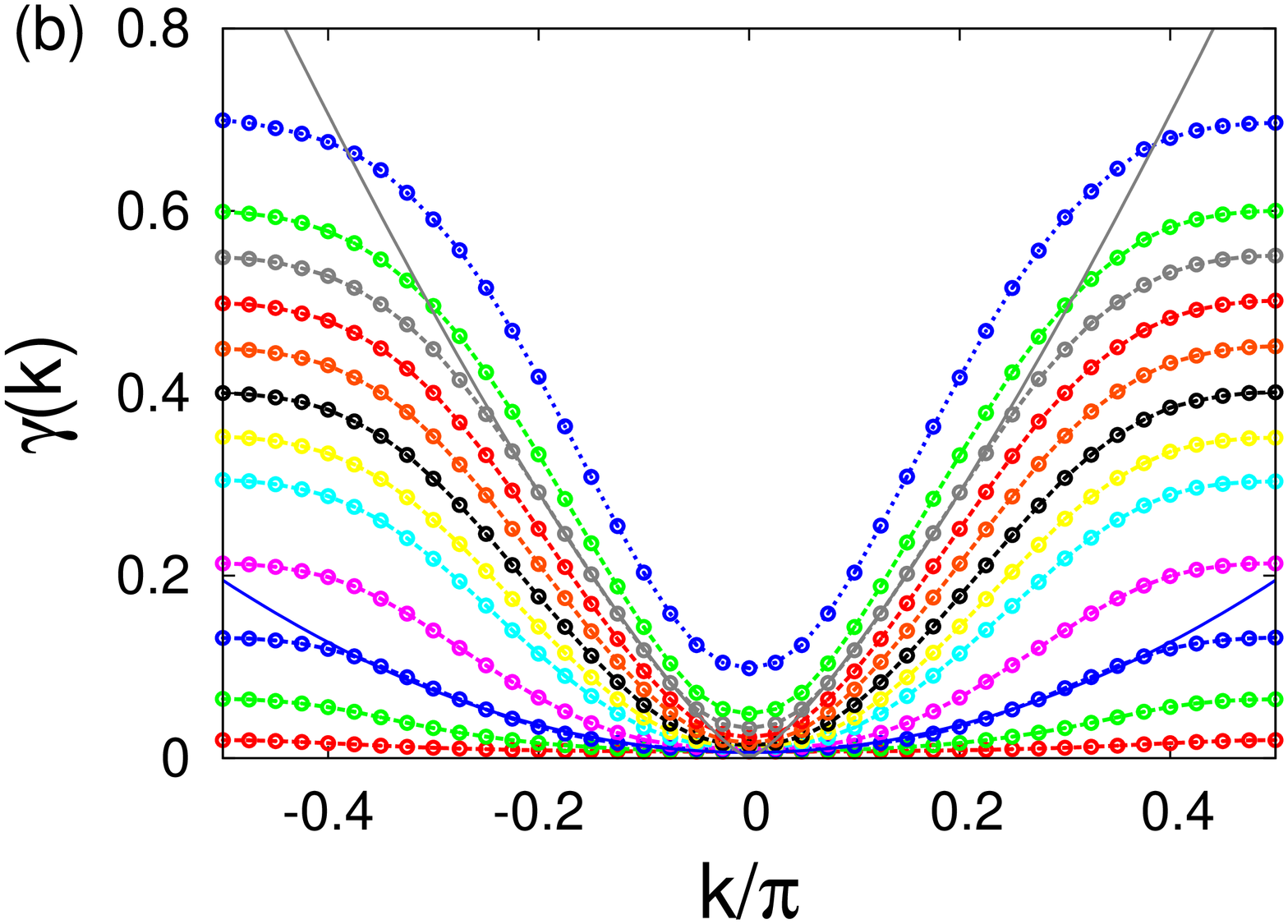}
\end{minipage}\hspace{0.5pc}
\begin{minipage}{.32\textwidth}
\includegraphics[width=0.75\linewidth,angle=-90]{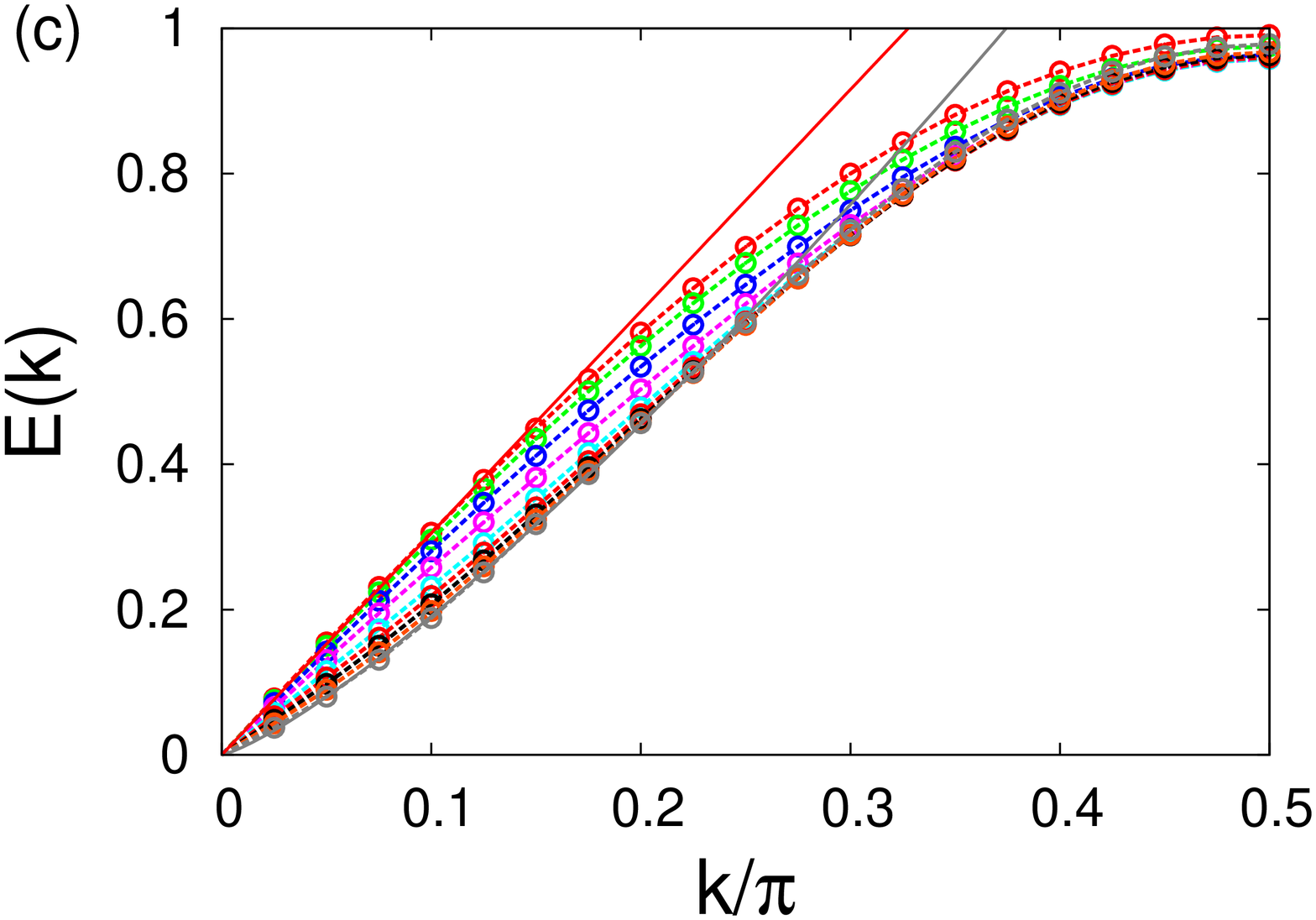}
\end{minipage}\hspace{0.5pc}
\caption{(color online) Momentum dependence of the single particle excitations for various Gaussian disorder strengths for $L=80$, $N_C=2^9$, and momentum $\bk=(k,0,0)$. (a) The residue, (b) spectral line width, (c) and dispersion for various disorder strengths. The solid grey lines are the power law fits at the AQCP ($W_c/t\approx 0.75$), and the solid blue [red] line is a fit of $\gamma(\bk)$ [$E(\bk)$]  to $\gamma(0)+a|\bk|^2$ [$v |\bk|$] at low momentum. }
\label{fig:Ek}
\end{figure*}

The existence of sharp excitations at the Weyl node in the SM regime would require
the damping rate $\gamma(\bk)$ to vanish faster than $E(\bk)\sim |\bk|$ at small $|\bk|$. 
 Focusing on the Weyl point, this question reduces to whether or not $\gamma(0)>0$.
In the SM regime the finite size DOS has low-$|E|$ peaks that are composed of perturbatively dressed Weyl states~\cite{Pixley1}. These produce 
sharp \emph{momentum resolved} Weyl peaks in $A({\bf k},\omega)$ at low-$|\omega|$ and low-$|\bk|$ as shown in Fig.~\ref{fig:Aw} (b).
Therefore, in addition to the intrinsic line-width that is non-zero in the thermodynamic limit there is also two additional spurious  contributions to the width of the peaks in the spectral function. These are due to (i) perturbative disorder broadening that vanishes in the large-$L$ limit and (ii) the finite KPM expansion order sets an extrinsic spectral width going as $\sim1/N_C$. The perturbative contributions to the energy go like $E \approx E_0 + E_1+E_2$, where $E_1 = \sum_{\br}V(\br)/L^3\sim W/L^{3/2}\times(\mathrm{random \,\,\, sign})$ and $E_2 \sim W^2/L^2 \times(\mathrm{random \,\,\, sign})$. By shifting the Gaussian distribution for each sample to satisfy $\sum_{\br}V(\br)=0$ we have set $E_1=0$ suppressing the leading finite size effect, and the perturbative contribution only appears at $W^2/L^2$ order.
To reach the intrinsic line width we converge the KPM data in both $N_C$ and $L$.
Our procedure is as follows, we first fix the system size to $L=80$ and increase $N_C$ until $\gamma(0)$ converges, we then increase $L$ for fixed $N_C=16384$, see Fig.~\ref{fig:k=0} (a) . This allows us to work in the SM regime ($W<W_c \approx 0.75t$), but due to the perturbative broadening we cannot go to arbitrarily low disorder strengths for the system sizes and expansion orders considered here. As shown in the inset of Fig.~\ref{fig:k=0} (a), we can converge $\gamma(0)$ in $L$ and $N_C$ down to $W=0.625t$. 

As shown in Fig.~\ref{fig:k=0} (b), we find that $\gamma(0)$
is exponentially small at weak disorder and is well fit by the rare region form (see Ref.~\cite{NandkishoreHuseSondhi})
\begin{eqnarray}
\log \gamma(0) \sim - (t/W)^2,
\label{eqn:rr}
\end{eqnarray}
which is one of our main results. 
We find excellent qualitative agreement [Fig.~\ref{fig:k=0} (b)] between the $T$-matrix calculation and the KPM data.  Therefore, we conclude that 
rare states produce excitations with an exponentially large (but not infinite at any finite disorder no matter how weak) quasiparticle lifetime, and there are no sharp quasiparticle excitations at the Weyl node for $W>0$.

We now turn to the residue of the Weyl node as a function of the disorder strength as shown in Fig.~\ref{fig:zk=0} (a). The $T$-matrix gives a weak correction and $Z(0)\approx 1$. This is in sharp contrast to a perturbative RG calculation~\cite{Pixley-2016B} that finds $Z(0)\sim (1-W/W_c)^{\nu(z-1- \eta)}$ signaling a non-analytic $G(\bk,\omega)$ at $W_c$. 
However, the numerically exact results are qualitatively in-between these two pictures, the quasiparticle residue is always non-zero but has a minimum near the AQCP (for Gaussian disorder~\cite{Pixley1} $W_c\approx0.75t$). At larger disorder $Z(0)$ is unaffected by passing through the Anderson localization transition [see inset of Fig.~\ref{fig:zk=0} (a)].
The binary disorder distribution suppresses the strength of avoidance, as shown in Fig.~\ref{fig:zk=0} (b) this leads to a smaller value of $Z(0)$ at the AQCP ($W_c\approx 0.86t$ for binary~\cite{Pixley2}).  Thus, the size of $Z(0)$ near $W_c$ is controlled by the strength of avoidance and the Green function is \emph{always} analytic near the quasiparticle peaks for $W>0$. This is an important new result.

\subsection{Disordered Weyl excitations}
We now come to the momentum dependence of the single particle excitations. Within the $T$-matrix calculation we find that the quasiparticle residue is weakly renormalized away from unity and develops momentum dependence at low energy, see Fig.~\ref{fig:Tmatrix_z} (a). The $T$-matrix calculation leads to a low-momentum damping rate going as
$
\gamma(\bk) \approx \gamma(0)+a(W) |\bk|^2,
$
where $\gamma(0)$ is given by Eq.~(\ref{eqn:rr}) and $a(W)$ is an increasing function of $W$, [See Fig.~\ref{fig:Tmatrix_z} (b).]

Turning to the KPM results shown in Fig.~\ref{fig:Ek}, at weak disorder and low momentum we find $E(\bk) \approx \pm v(W) |\bk|$, $Z(\bk) \approx \mathrm{const}$, and the damping is well described by 
$
\gamma(\bk) \approx \gamma(0) + \tilde{a}(W) |\bk|^2
$
in good agreement with the $T$-matrix results.
 For larger disorder strengths, approaching the AQCP ($W\approx W_c$) we find a momentum regime $k^* < |\bk| < \Lambda/v$ where the single particle excitations develop clear powers going like $E(\bk) \sim \pm|\bk|^{1.3}$, $\gamma(\bk) \sim |\bk|^{1.3}$, and $Z(\bk)\sim |\bk|^{0.4}$. The low momentum cross over scale $k^*$ is determined by when $\gamma(\bk)$ is comparable to $E(\bk)$.  Our results are consistent with the single particle excitations developing non-trivial power laws in the QC regime: $E(\bk)\sim \pm |\bk|^z$, $\gamma(\bk) \sim |\bk|^{d-z}$, and $Z(\bk)\sim |\bk|^{z-1- \eta}$, where $z=1.5$ and $\eta=0.125$ (within a modified RG scheme~\cite{Goswami-2016}).  However, due to the strong avoidance and the resulting limited scaling regime, 
 the momentum power laws provide a poor estimate of $z(=1.5\pm0.04$, Ref.~\onlinecite{Pixley2}) and $\eta$.
 
 \subsection{Off-shell rare region contribution}
So far we have been concerned with the properties of the Green function in the vicinity of the quasiparticle peaks and the renormalization of the ``on-shell'' single particle properties, i.e. the renormalization of the pole in $G(\bk,\omega)$. In this subsection we consider the low energy``off-shell'' contributions to the spectral function far away (in energy) from the quasiparticle peaks. 
For the DOS by using a twist, we are able to move the Weyl states away from zero energy to separate the perturbative and rare contributions, this is not possible when we consider the momentum resolved spectral function since a twist will just move the location of the Weyl cone.

\begin{figure}[h!]
\centering
\begin{minipage}{.45\textwidth}
\includegraphics[width=0.75\linewidth, angle=-90]{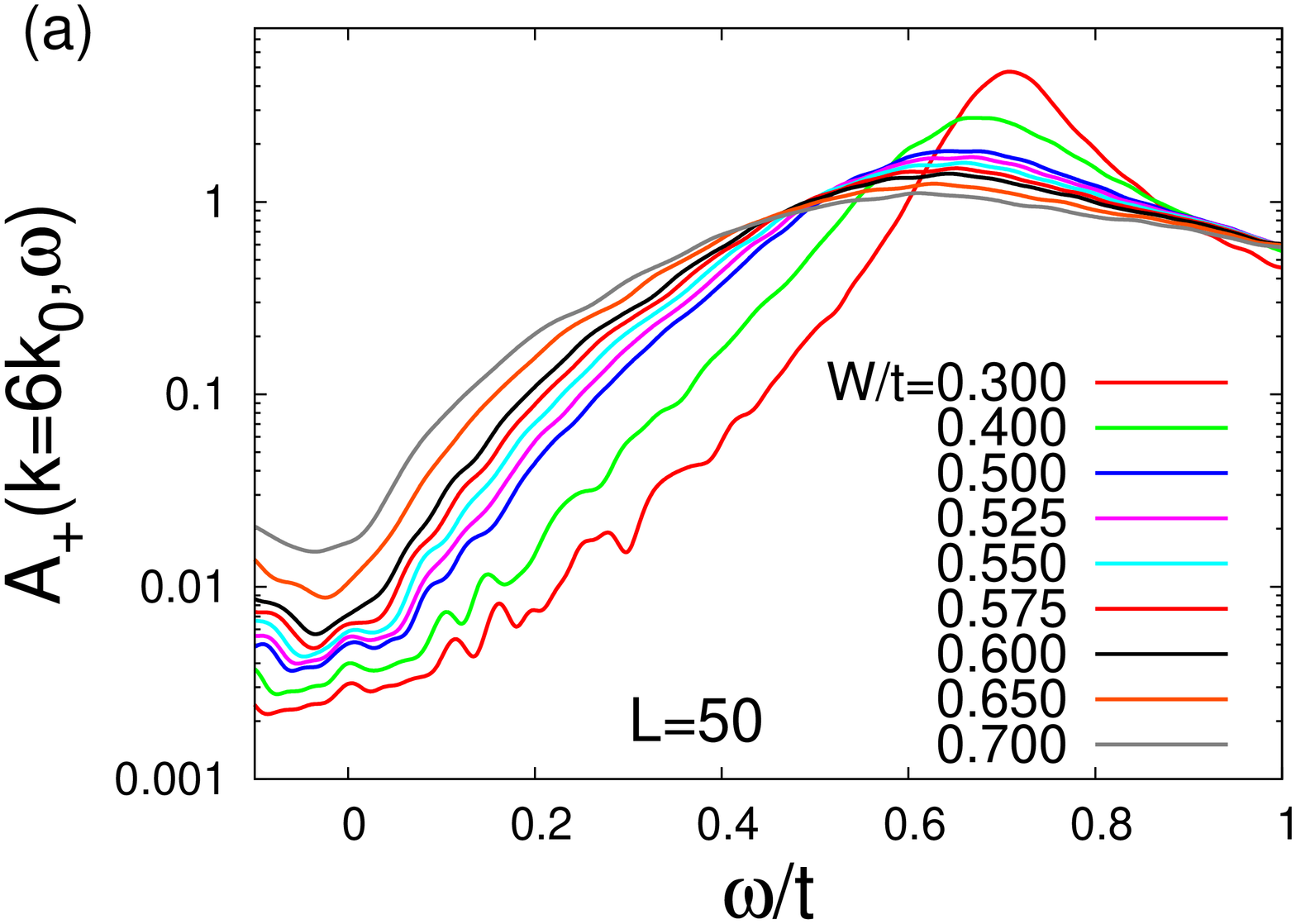}
\end{minipage}\hspace{0.5pc}
\centering
\begin{minipage}{.45\textwidth}
\includegraphics[width=0.75\linewidth, angle=-90]{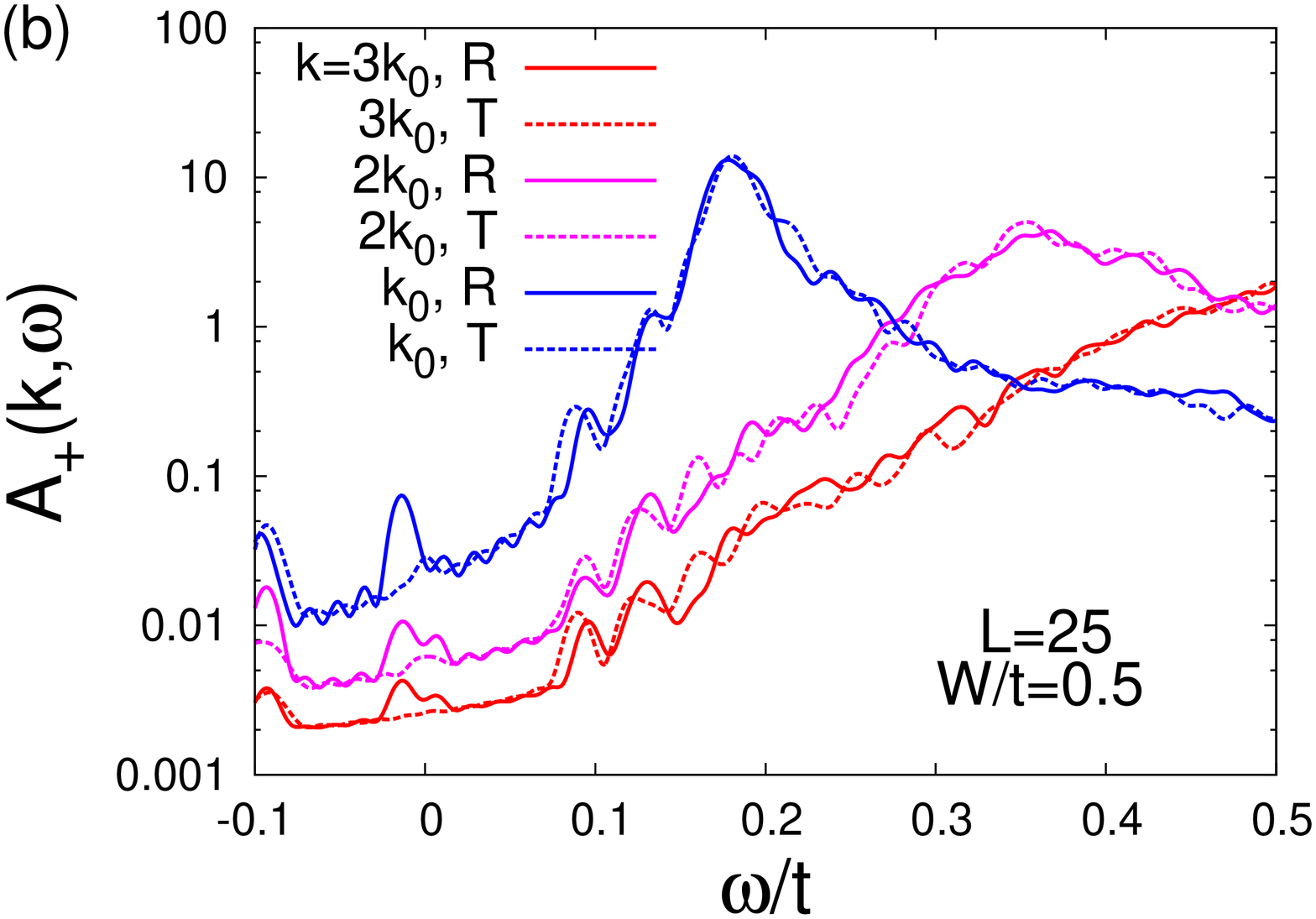}
\end{minipage}
\caption{(color online) Off-shell rare region contribution to the spectral function. (a) The average spectral function for $L=50$, $N_C=512$, $k=6k_0 = 6(2\pi/L)$, and various disorder strengths. At weak disorder the spectral function is peaked near $k$ but there is a clear peak near $\omega=0$. (b) Spectral function for a single sample that is either rare (R) or typical (T) with $L=25$, $W/t=0.5$, $N_C=1024$, and various momentum ($k_0=2\pi/L$). We find the rare sample has a clear peak near zero energy at all momentum we consider here that is absent in the typical sample.}
\label{fig:rareAkw}
\end{figure}

As shown in Refs.~\onlinecite{NandkishoreHuseSondhi,Pixley1}, rare low energy eigenstates give rise to an exponentially small but non-zero DOS for an arbitrarily small disorder strength $\rho(0) \sim \exp(-a/W^2)$. Whereas, typical perturbatively dressed Weyl states produce peaks in the finite-size DOS, that become momentum resolved Weyl peaks in $A(\bk,\omega)$ [see Figs.~\ref{fig:Aw} (b), (c), and (d)]. It is therefore an interesting and natural question to ask how will rare states contribute to the spectral function? These rare eigenstates are quasi-localized about the site with a very large disorder strength and fall off like $1/r^2$ at short distances. As a result we expect these wave functions to be broad in momentum space and they can in principle have a non-zero overlap with almost any clean plane wave like state. To see this we will consider the ``off-shell'' contributions of states near $\omega=0$ for $k>0$.  As shown in Fig.~\ref{fig:rareAkw} (a) the average spectral function for $k =6 (2\pi)/L$ is peaked at $k$ for weak disorder and in addition there is a clear peak that is centered about $\omega=0$ with an amplitude that is several orders of magnitude below the on-shell peak at $k$. To understand the origin of the peak at $\omega=0$ we consider $A(\bk,\omega)$ for a single disorder sample in Fig.~\ref{fig:rareAkw} and take two different disorder configurations -- one rare and one typical -- where we know a priori one sample produces a rare state near low-$|\omega|$ and the other sample does not~\cite{Pixley1}. As shown In Fig.~\ref{fig:rareAkw} (b), we find that for the rare disorder sample there is a peak near $\omega=0$ for each $k$ shown, whereas the typical sample does not have a peak near zero energy. Therefore, we conclude rare states produce an off-shell contribution near $\omega=0$ to $A(\bk,\omega)$ that produces a peak in the finite-size average spectral function.

\subsection{Anomalous dimension of AQCP}
We now consider the quantum critical scaling of $G'$ and compute the anomalous dimension ($\eta$) of the Weyl field at the AQCP, 
which is defined as 
\begin{equation}
G(\mathbf{k},0) \sim 
\hat{k} \cdot \boldsymbol \sigma/k^{1+\eta}.
\end{equation}
Using the $d=2+\epsilon$ expansion scheme at one loop order one finds $\eta=0$ (see Refs.~\onlinecite{Goswami-2011, Syzranov-2015L, Syzranov-2015, Pixley-2016B}) and is unchanged from $W=0$. But, a modified RG scheme~\cite{Goswami-2016} at one loop leads to 
\begin{equation}
G(\mathbf{k}, \omega) \sim \frac{e^{-(z-1-\eta) l}}{\omega+i\delta-v \mathbf{k}\cdot \boldsymbol \sigma e^{(1-z)l}},
\end{equation}
with $z=1+\epsilon/2$ and $\eta=\epsilon/8$.  
In the critical regime $e^l \sim 1/|\mathbf{k}|$ and $G'(\mathbf{k},0) \sim 1/|\mathbf{k}|^{1+\eta}$
with $\eta=0.125$, $\nu=1$ and $z=1.5$ (after setting $\epsilon=1$). 

\begin{figure}[h!]
\centering
\includegraphics[width=0.75\linewidth,angle=-90]{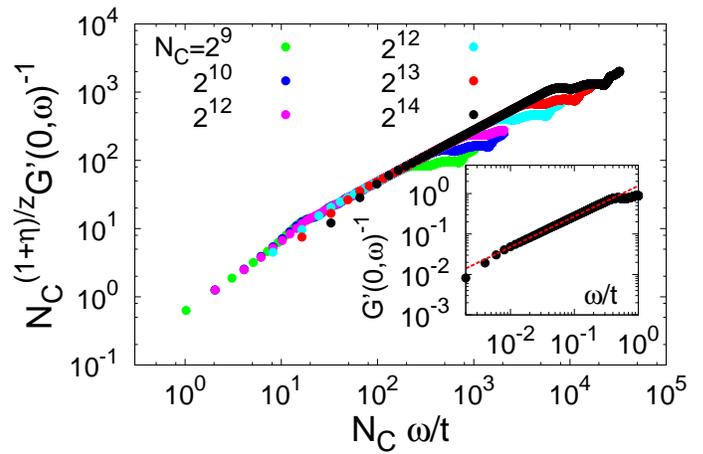}
\caption{(color online) Computing the anomalous dimension $\eta$ at the AQCP ($W_c/t=0.86\pm0.01$) for binary disorder and a system size of $L=80$: (Inset) Fit of $1/G'(k=0,\omega)$ to a power law in the quantum critical regime yields $\eta = 0.13 \pm0.04$. Data collapse in the KPM expansion order $N_C$, showing excellent single parameter scaling for over two decades of $N_C\omega$.}
\label{fig:g'}
\end{figure}

Turning to the KPM, due to the limited quantum critical range in momentum in Fig.~\ref{fig:Ek}, we focus on the corresponding scaling in energy 
$
G'(\bk=0,\omega)\sim 1/\omega^{(1+\eta)/z}.
$
Here, we use binary disorder to get the largest critical scaling regime before the avoidance rounds it out~\cite{Pixley2}.
As seen in the inset of Fig.~\ref{fig:g'} using binary disorder, we have a clear power law fit for about a decade and a half, which yields $(1+\eta)/z=0.75\pm0.01$, thus $\eta=0.13\pm0.04$ (using $z=1.5\pm0.04$, Ref.~\onlinecite{Pixley2}). 
Note that at the lowest energies the data falls off of this power law as the model crosses over to the DM regime.
This is in excellent agreement with the modified RG scheme. Due to the KPM broadening, the finite expansion order acts like an effective inverse energy scale that can round out the AQCP~\cite{Pixley2}. As a result, we expect that in the quantum critical regime the Green function obeys single parameter scaling 
\begin{equation}
G'(0,\omega)^{-1} \sim N_C^{-(1+\eta)/z}f(\omega N_C),
\end{equation}
where $f(x)$ is a scaling function. As shown in  Fig.~\ref{fig:g'}, we find excellent data collapse for over two decades in $N_C\omega$. 

\section{Conclusions}
\label{sec:conclusions}
We have investigated the single particle Green function in disordered Weyl semimetals. We have employed various techniques including the $T$-matrix, the renormalization group, and the numerically exact kernel polynomial method.
Our results can be directly tested in STM and ARPES experiments in compounds dominated by neutral defects (e.g. vacancies and interstitials) to avoid ``doping'' the Weyl cone due to the screening of charged impurities~\cite{Skinner,Graphenereview2}. 
In general, approaching the Dirac-Weyl point is a formidable experimental challenge even in a system as well-studied as graphene because of inhomogeneous density puddles induced by random charged disorder invariably present in the material~\cite{Adam-2007}, but a combination of materials improvement with little residual  Coulomb disorder and not too low a temperature~\cite{DasSarma-2013} may very well lead to a verification of our predictions.
At weak disorder, our results predict a spectral line-width that scales like $|\bk|^2$ at moderate $|\bk|$ but saturates to a non-zero constant at (exponentially) small $|\bk|$. 
Detecting the exponentially small $\gamma(0)$ at finite temperature will be challenging on very clean samples, but the broadening of the line shapes with increasing $|\bk|$ should be accessible.
At moderate disorder strengths, elastic scattering induces line shapes that are broad and asymmetric with strongly renormalized single particle properties due to the avoided quantum critical point. This will produce a measurable effect by reducing the observed Fermi velocity.

We have provided the first numerical estimate of the anomalous dimension of the avoided quantum critical point, with $\eta=0.13 \pm0.04$. It is interesting that the two different RG schemes we have discussed yield results that differ in their estimate of $\eta$. A priori it is in no way obvious which RG procedure yields more accurate results. The `conventional' $2+\epsilon$ expansion yields a result that is unchanged from the clean limit ($\eta=0$). In contrast, the use of a spatially correlated disorder distribution in Eq.~(\ref{eqn:pot}) has lead to $\eta=1/8$, which is in excellent agreement with the numerics. Thus the combined RG and numerical analysis allows us to conclude that the spatially correlated disorder procedure is in fact more accurate then the conventional RG treatment. 

We have demonstrated that quasiparticle excitations are sharp in the semimetal regime and marginally broadened in the quantum critical regime. However, at sufficiently low energy and weak disorder, non-perturbative effects of disorder dominate and excitations at the Weyl node always have a finite lifetime.  Our numerically exact results have established that the disorder averaged single particle Green function remains analytic near the quasiparticle peaks, but obeys single parameter scaling in the cross over regime at finite energy. Lastly, we have shown how rare states contribute to both the on-shell and off-shell part of the spectral function.
Our work establishes definitively how quasiparticle spectral properties in disordered Weyl fluids directly reflect both the avoided criticality and the rare region effects in subtle, but well-defined, manners.

\noindent\emph{Note Added}--- During the review of our work we became aware of the publication Phys. Rev. Lett. 118, 146401 (2017). Our work disagrees with this publication and our findings related to non-perturbative effects of disorder and avoided critical scaling invalidate their conclusions. 

\acknowledgements{\emph{Acknowledgements}: We acknowledge useful conversations with John Chalker and Victor Gurarie. This work is partially supported by JQI-NSF-PFC, LPS-MPO-CMTC, and Microsoft Q (J.H.P., P.G., and S.D.S.); by a Simons Investigator award
from the Simons Foundation and NSF grant no. DMR-1001240 (Y.-Z.C. and L.R.); and by  the KITP under Grant No.
NSF PHY-1125915 (R.N. and L.R.). We acknowledge the University of Maryland supercomputing resources (http://www.it.umd.edu/hpcc) made available in conducting the research reported in this paper. We thank the KITP for its hospitality during our stay, when part of this work was completed (R.N. and L.R.). J.H.P. also acknowledges the hospitality of the University of Colorado Boulder.
}

\bibliography{DirtyWeyl_bib}
\end{document}